\documentclass[AER]{AEA}

\usepackage[sectionbib,round]{natbib}
\usepackage{graphicx}
\usepackage{multicol}
\usepackage{caption}
\usepackage{dcolumn,booktabs}
\newcolumntype{d}[1]{D{.}{.}{#1}}
\usepackage{caption}
\usepackage{longtable}

\usepackage[dvipsnames]{xcolor}
\usepackage{pgfplots}
\pgfplotsset{compat = newest}
\usetikzlibrary{positioning, arrows.meta}
\usepgfplotslibrary{fillbetween}
\usepackage{amsmath}

\usepackage{mathrsfs}
\usepackage{amsmath}        
\usepackage{amssymb,bbold}  

\draftSpacing{1.5}

\begin{document}
\title{Canceled: A New Reliability Incentive for Energy-Only Electricity Markets}
\shortTitle{Canceled: A New Reliability Incentive}
\author{Devin Mounts and Robin M. Cross\thanks{Mounts: Department of Applied Economics; Oregon State University, 213 Ballard Hall, Corvallis, OR 97331 and Portland General Electric, 121 SW Salmon St. Portland, OR, 97204. mountsd@oregonstate.edu. Cross: Dept. of Applied Economics; Oregon State University, 221B Ballard Hall, Corvallis, OR 97331, robin.cross@oregonstate.edu. Acknowledgments: We extend appreciation to Severine Borenstein for his market participant response suggestions provided as part of the Energy, Resources and Environmental Economics Student Seminar at U.C. Berkeley, which are incorporated as Section V of this paper, Carey King from the University of Texas Energy Institute and Joshua D. Rhodes from the University of Texas Webber Energy Group for providing critical background in ERCOT's market protocols and data, Ryan King, Pamela Shaw, Jonas Kersulis, Jian Chen, Zhengguo Chu, Haibo You, and Fred Adadjo from the ERCOT Market Design Group for their review of market structure assumptions and findings and William Hogan from Harvard's Kennedy School for his clarifications on the ERCOT's original incentive design and dynamics. We declare no financial interests related to the research within this paper. Disclaimer: This research was prepared by Devin Mounts in his personal capacity. The opinions in this article are the author's own and do not reflect the view of Portland General Electric.}}
\date{\today}
\pubMonth{Month}
\pubYear{Year}
\pubVolume{Vol}
\pubIssue{Issue}
\JEL{H23, L11, L94, Q41}
\Keywords{energy-only markets, ERCOT, orthogonalization, price incentives, reliability}

\begin{abstract}
This paper considers the reliability problem in energy-only markets. Following widespread blackouts in 2011, Texas introduced a reliability price incentive to attract two GW of net additional natural gas-generating capacity. The incentive is unusual because energy buyers pay the incentive directly to producers in a real-time spot market. The program has created \$13 billion in direct payments to generators annually since 2015 and is now being implemented or considered in several major energy markets in the US and abroad. We assess the incentive's impact on the Texas market from three perspectives: First, we derive the incentive's equilibrium effect on the electricity market price in a monopolistic market from first principles using a standard partial equilibrium economic model. We then empirically test whether the incentive encouraged net entry into the market or the generating applicant pool, controlling for market and climatic conditions using monthly capacity data. Finally, we look for direct evidence of an incentive response among active traders using real-time market trading data. The three approaches suggest buyers and producers cancel out the incentive, and the price-only program does not encourage new generation capacity to enter the market.
\end{abstract}

\maketitle

Under intense pressure to improve grid reliability following the state-wide blackouts of 2011, Texas enacted its first producer reliability incentive program in 2014 (the Incentive). The Incentive pays producers a price premium that increases as scarcity conditions in the market worsen. It is unique because buyers pay the price premium directly to producers as a lump sum per megawatt in real time. The Electric Reliability Council of Texas (ERCOT) anticipated the Incentive would generate an additional \$1.12B in annual energy profits necessary to replace 8,000 MW of gas generation capacity expected to retire due to tighter emissions standards and to attract 2,000 MW of net new capacity (\citealp{ercotBackCast}, \citealp{ercotForecastExits}). Since ERCOT's implementation, the program has generated an average of \$13.3 billion in incentive payments to generators annually, and several major U.S. wholesale power markets are considering or implementing a similar buyer-paid incentive approach. However, only 407 MW of net new capacity has entered the market.\footnote{Scarcity pricing models are being implemented by the Southwest Power Pool, New York Independent System Operator, Independent System Operator New England, Midcontinent Independent System Operator, and the Pennsylvania-New Jersey-Maryland Interconnection (PJM) \citep{hoganERCOT, pjmResources}].} 

\citet{cramptonStoft} and \citet{FABRA2018323} questioned the potential impact of short-run payments on long-run capacity growth. Texas reliability is again under increased scrutiny after ERCOT was forced to conduct the largest rolling blackout in US history in 2021.\footnote{\citealp[pg. ~9]{fercStormReview}} In this research, we analyze whether the simultaneous tax and subsidy policy canceled itself out or promoted additional natural gas capacity as intended.\textsuperscript{,}\footnote{Net growth is a result of forecast retirements \citep{ercotForecastExits}, additional energy payments resulting from policy \citep{ercotBackCast}, and estimated cost of entry for natural gas generators \citep*{coneEstimates}}

The reliability of an electrical system is not only a measure of total grid capacity but also the availability of that capacity to meet demand when required. This is made more difficult as the amount of renewable capacity in a market increases since the reliability contribution of a dispatchable resource, one that can be turned on and off, such as natural gas, is not equal to that of an intermittent resource which produces when the weather conditions allow, such as wind and solar. 

\begin{figure}
    \centering  
    \includegraphics[height=6.5cm]{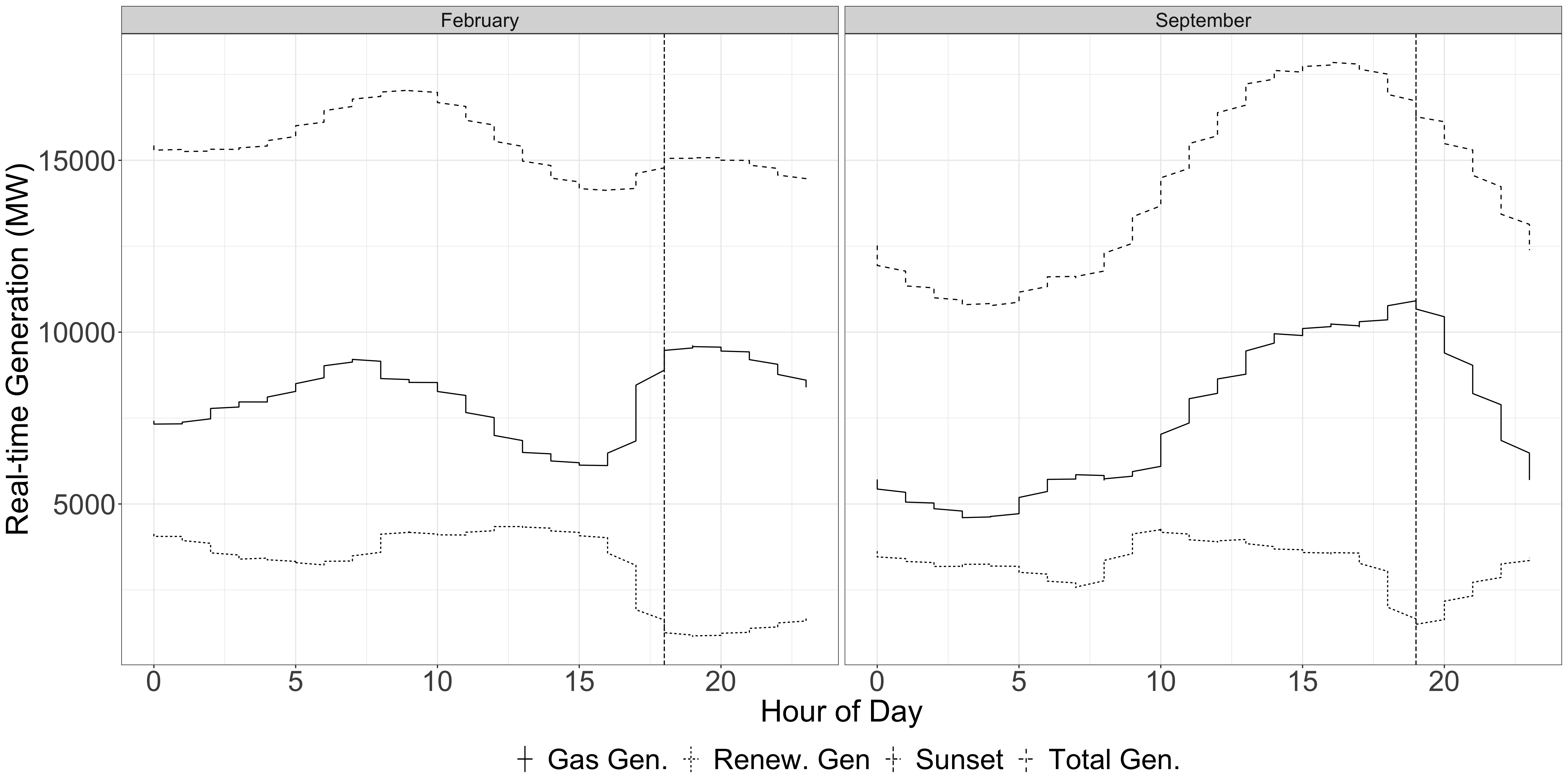}
    \caption{Generation by Resource Type for Peak Day of Month in ERCOT}
    \label{fig:enter-label1}
\end{figure}

Figure \ref{fig:enter-label1} highlights the complementary relationship between renewables and natural gas generation and the role of readily dispatchable generation from sources such as natural gas in providing reliability. Here, we see natural gas and renewable generation during two days in 2022: February (left) and September (right). In February, total generation peaked in the morning and again in the evening due to heating demand. As the sun set, generation from renewables fell by approximately 2,000 MWh, requiring a 3,000 MWh increase from natural gas to meet evening peak demand. There was a more prominent single peak in September, driven by cooling demand. Again, as the sun set, renewable generation subsided, and natural gas generators increased production to compensate. The mirrored natural gas and renewable generation curves illustrate each resource's complementary contribution toward reliability.
\begin{figure}
    \centering  
    \includegraphics[height=8cm]{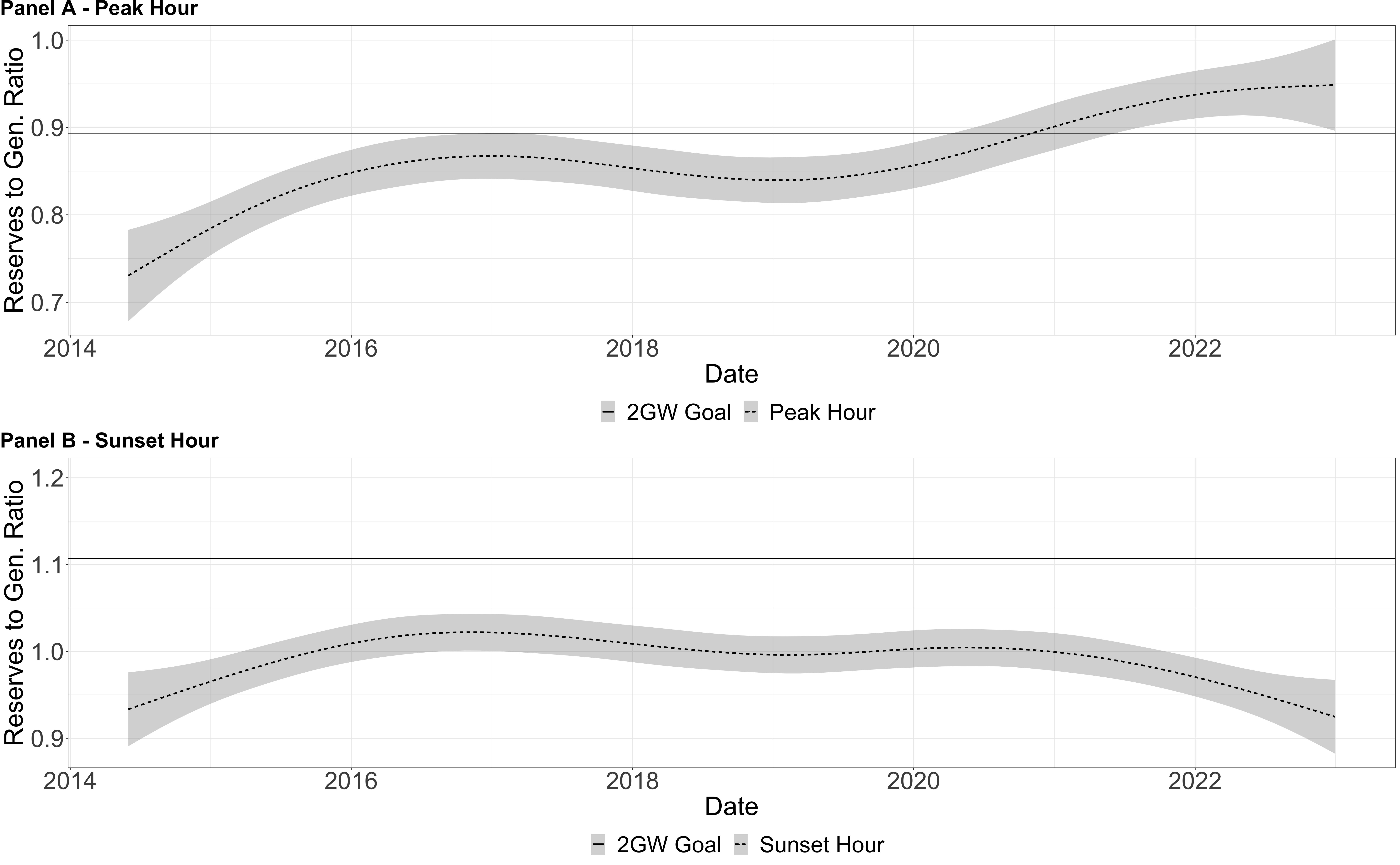}
    \caption{Daily Ratio of Reserves to Demand at Peak (A) and Sunset (B) Hours}
    \label{fig:enter-label2}
\end{figure}

The challenge of measuring reliability in a market with large quantities of renewable generation is depicted in Figure 2. Panel A shows the ratio of reserves to generation in ERCOT at the peak hour of each day from 2014, when the Incentive was implemented, through 2022.\footnote{Daily reserve point estimates are smoothed using a generalized additive model.} The dotted line displays the upward trend in this ratio at the peak hour of the day, which averages 4:00 pm in the Summer and 12:30 pm in the Winter. The horizontal line indicates the policy's 2,000 MW net growth goal. Reserves surpass this goal in 2021, suggesting that the policy was successful. Panel B, in contrast, shows the same ratio at the sunset hour, on average 8:15 pm in the Summer and 5:30 pm in the Winter. The sunset hour ratio shows a flat-to-declining trend from 2016, never reaching the additional 2,000 MW of dispatchable reserves marked by the horizontal line. As suggested in Figure 1, generation from renewables declines at or near sunset, suggesting Panel B more accurately reflects the changes in grid reliability over time. However, this illustration is not scientific because it does not measure how much capacity entered the market and whether that entry was a result of the policy or other market changes relevant to the firm's entry decision, areas we will focus on in this study. 

The implications of our research go beyond whether ERCOT's approach has worked for Texas. Many large US energy markets are undergoing a rapid energy transition as 75\% of US coal generation is estimated to retire by 2040 to curb climate change \citep{forecastCoal}. The US's largest wholesale power market, Pennsylvania-New Jersey-Maryland Interconnection (PJM), announced a projected 21\% reduction in capacity by 2030 as 40,000 MW of fossil-fueled generation resources are expected to retire due to environmental and carbon emissions regulations and industry goals for clean energy \citep{pjmResources}. The PJM recently implemented a price-based policy approach, similar to ERCOT's, in addition to a secondary market to promote long-term capacity. Other markets are adopting ERCOT's approach to reliability, even though the effect of this policy is still unknown.

Three features of Texas' ERCOT market make it a valuable setting for testing the effect of this policy design. First, the Incentive is the only reliability policy in ERCOT, where other markets are incorporating this approach in addition to creating markets specifically to procure additional reserve generation capacity. Second, ERCOT is not integrated with other US electricity markets. This lack of integration results in ERCOT relying solely on generation within its own market, as it cannot import or export large quantities of power. Finally, supply changes in ERCOT's generation portfolio reflect transitions underway in other US markets, as the retirement of coal facilities and the substantial growth of intermittent renewable resources cause dispatchable natural gas to shoulder the burden of reliability. These factors create a unique environment where we can test the impact of the Incentive on capacity growth in the ERCOT market over the past seven years.

Our research extends the literature in three areas. First, we extend the study of subsidies and taxation, explored by \citeauthor{Jenkin} 's (\citeyear{Jenkin}) seminal work on physical neutrality and extended to imperfectly competitive markets by \citeauthor{Weyl2013} (\citeyear{Weyl2013}).\footnote{We refer to the Incentive here as a simultaneous buyer tax and producer subsidy to frame it in terms of the partial equilibrium theory results of \citeauthor{Jenkin} (\citeyear{Jenkin}) and \citeauthor{Weyl2013} (\citeyear{Weyl2013}), but alternative terms such as surcharge \citep{EconomidesSurcharge}, user fee \citep{UserFees}, and incentive \citep{Incentives} are also relevant.} We are the first to formally identify and analyze a firm subsidy paid directly by the consumer in a competitive and monopolistic market. We develop a theoretical model to formalize the Incentive and demonstrate how it influences behavior among profit-maximizing market participants. Second, we extend the literature on energy-only markets, price caps, and reliability through our empirical analysis of the Incentive's impact on entry into and exit from the ERCOT generating pool using a recursive structural model. Earlier work by \citet{Cabral1991} suggested price caps reduce capital investment in monopolistic markets. \citet{hoganIntroduce} then proposed a price-only incentive program to counter under-investment, presented as a solution to reliability in ERCOT \citep{hoganERCOT}. \citet{cramptonStoft} critiqued Hogan's incentive design, suggesting short-run price incentives alone are too erratic to support a reliable surplus of supply, and proposed a separate market for reserve capacity to provide a longer-term incentive. The separate-market design was simulated for ERCOT by \citet{BAJOBUENESTADO2017272}, who found increased reliability and prices. Our analysis suggests that the Incentive does not increase firm profits, does not encourage additional generating capacity in ERCOT, and has an unintended net negative effect on market prices. Finally, our findings add to the ongoing examination of reliability and market power within the ERCOT energy market \citep{hortacsu, dyer2011, coneEstimates}.

Although we critique the direct consumer payment and price-only aspects of ERCOT's Incentive design, we do not suggest that other approaches to reliability, whether regulated reserve margins or capacity markets, might not improve reliability in ERCOT. Finally, we do not claim that future reliability in the market is dependent only on natural gas but rather that ERCOT has failed to encourage entry from such dispatchable sources through the Incentive.

The rest of the paper is organized as follows: Section I provides background on ERCOT's Incentive and describes the data. Section II presents the partial equilibrium framework and shows that the Incentive does not influence firm profits. Section III tests the framework empirically, introducing a structural model to capture the firm's entry decision and remove multicollinearity and then testing the Incentive's impact, controlling for energy prices, weather events, and other market factors. Section IV tests for evidence traders lower bids and asks in response to the Incentive level, as predicted in the partial equilibrium model, using real-time market data, and Section V concludes.

\section{Background and Data}
\subsection{Background and Relevant Literature}

ERCOT implemented its Incentive in June 2014, as initially proposed by \citet{hoganERCOT}, to offset the potential loss of revenue from a newly imposed price cap \citep{cramptonStoft} and to improve reliability. From Hogan's analysis, ERCOT estimated the Incentive would have increased the profitability of a gas-fueled generator by 17\% during 2011, or an additional \$21,794 to \$125,001 of profit during the year per generator (\citealp[pp.~10--11]{ercotBackCast}). For comparison, the estimated cost for a one MW gas generator to enter the market in that same year was \$116,000 (\citealp[p.~48]{coneEstimates}).

Since its implementation, we have found no studies that analyze the causal influence of the Incentive on capacity growth. ERCOT's internal estimates suggest a correlation between a higher Incentive and the quantity of reserves in the market over time. Recent adjustments to the Incentive design are reported to have improved reserve capacity, as illustrated in Figure \ref{{fig:enter-label2}} panel A. However, their analysis does not control for economic, climatic, or market factors that influence capacity growth \citep{ercotReview2022}. 

Several studies have raised concerns over the theoretical justification for the Incentive design initially outlined by \cite{hoganIntroduce}. \cite{cramptonStoft} agree prices are one aspect of an efficient reliability incentive but argue a price-only incentive does not provide a stable signal of long-term cost recovery that investors require to enter the market and propose instead a separate Capacity Market to pay producers to maintain offline reserve capacity. \cite{BAJOBUENESTADO2017272} extends the Capacity Market approach by simulating the impact of offline capacity payments in the ERCOT market setting and finds these payments would increase reliability in the market and raise the cost to the consumer. In addition to the literature, we note one concern an anonymous ERCOT staff member raised during the Incentive design process pertinent to our examination: that producers may reduce bids to offset the Incentive \citep{ercotDesignReview}. This concern provides the intuition behind our theoretical and empirical findings.

\subsection{Data}

Two primary data sets are used for our analysis. First, monthly inventories of proposed, generating, postponed, and retired generation units are provided by the US Energy Information Administration (EIA) Form 860M from July 2015 through December 2022. The data provides generator-level information on capacity, status, fuel source, plant, operator, first operation date, retirement date, and geo-coordinates for all US plants with more than one MW of capacity. We narrowed this sample to 2,249 verifiable generators within ERCOT during the analysis period.\footnote{See the appendix for details about sample construction.}

Table \ref{A} summarizes active ERCOT generators by count and capacity in our sample, grouped into three fuel source categories: Natural Gas, Renewables (Wind and Solar), and Other (Coal, Nuclear, Hydro, etc.). Using our monthly observations, we measure the entry and exit of capacity for these three categories and the generators present in the initial period (July 2015) and the final period (December 2022). From this information, we can see that the total capacity in ERCOT increased from 97,654 MW to 125,860 MW for a net growth of 28,206 MW (entry minus exit), or 29\% from 2015. 407 MW of this growth resulted from the entry of 367 natural gas generator units. These 367 new natural gas generators each have 15 MW of capacity on average, 94 MW smaller than the average retired. Renewable capacity increased by 31,272 MW, growing from 15\% of total market share in 2015 to 36\% by the end of 2022 and accounting for 80\% of entry in ERCOT. Finally, 5,718 MW from Other generators exited the market, primarily Coal, 60\% of the total market exit.

\begin{table}[!htbp] \centering 
  \caption{Form 860M: ERCOT Active Generators - Entry and Exit} 
  \label{A} 
\small 
\begin{tabular}{@{\extracolsep{5pt}} llrrrr} 
\\[-1.8ex]\hline 
\hline \\[-1.8ex] 
Group & Measure & NG & Renewables & Other & Total \\ 
\hline \\[-1.8ex] 
Initial Period & Generating Units & 500 & 112 & 476 & 1,088 \\ 
 & Share Total Units & ( 0.46 ) & ( 0.10 ) & ( 0.44 ) & ( 1 ) \\ 
 & Mean MW & 115.09 & 130.08 & 53.89 & 89.92 \\ 
 & Total MW & 57,542.60 & 14,568.50 & 25,543.00 & 97,654.10 \\ 
 & Share Total MW & ( 0.59 ) & ( 0.15 ) & ( 0.26 ) & ( 1 ) \\ 
Entrants & Generating Units & 367 & 205 & 63 & 635 \\ 
 & Share Total Units & ( 0.58 ) & ( 0.32 ) & ( 0.10 ) & ( 1 ) \\ 
 & Mean MW & 15.43 & 154.44 & 35.65 & 62.17 \\ 
 & Total MW & 5,664.40 & 31,506.00 & 2,246.20 & 39,416.60 \\ 
 & Share Total MW & ( 0.14 ) & ( 0.80 ) & ( 0.06 ) & ( 1 ) \\ 
Exits & Generating Units & 48 & 3 & 76 & 127 \\ 
 & Share Total Units & ( 0.38 ) & ( 0.02 ) & ( 0.60 ) & ( 1 ) \\ 
 & Mean MW & 109.54 & 78.00 & 75.25 & 88.27 \\ 
 & Total MW & 5,257.90 & 234.00 & 5,718.70 & 11,210.60 \\ 
 & Share Total MW & ( 0.47 ) & ( 0.02 ) & ( 0.51 ) & ( 1 ) \\ 
Final Period & Generating Units & 819 & 314 & 463 & 1,596 \\ 
 & Share Total Units & ( 0.51 ) & ( 0.2 ) & ( 0.29 ) & ( 1 ) \\ 
 & Mean MW & 70.76 & 146.46 & 47.88 & 79.01 \\ 
 & Total MW & 57,949.10 & 45,840.50 & 22,070.50 & 125,860.10 \\ 
 & Share Total MW & ( 0.46 ) & ( 0.36 ) & ( 0.18 ) & ( 1 ) \\ 
\hline \\[-1.8ex] 
\multicolumn{6}{l}{Values in parenthesis denote share of total} \\ 
\end{tabular} 
\end{table} 

Generator entry is identified for units recorded with either "Operating" or "Stand-by/Backup" status during a month more recent than our initial observation period. Generator exits are identified when the generator's status is no longer recorded as "Operating," "Stand-by," or "Out of Service, but expected to return to service in the next calendar year," i.e., it is recorded as out of service or retired. By grouping available capacity by the three fuel source categories, we measure the generating capacity for every month in the analysis period. 

While we can identify when a generator enters operation or retires with dates recorded in the data, we cannot confidently determine when a unit enters or exits any of the six applicant pool stages that precede operation. There are six applicant statuses before operation begins, spanning from "Planned for installation, but regulatory approvals not initiated" to "Construction Complete, but not yet in commercial operation." A total of 93\% of our sample's megawatts from natural gas generators appear in at least one of the six applicant pool phases before generation. However, only 2\% of the megawatts from these generators appear in the first applicant pool phase. Considering our inability to track entry into the applicant pool, we focus our empirical analysis on changes to generating capacity actively operating in the market. Given that most units appear in at least one phase of the applicant pool before operating, we consider the sum of all units in any stage of the applicant pool statuses as a representative measure of the applicant pool in any given month for the same three fuel source categories, Natural Gas, Renewables and Other. 

The second data set comes from ERCOT's Real-time Prices Report and the Historical Real-Time ORDC and Reliability Deployment Price Adders and Reserves Report, which provides market-level price and generation at fifteen-minute intervals and supply reserves and the Incentive at five-minute intervals. We use this data in both of our empirical analyses. First, in Strategy 2, we calculate monthly statistics from this market data and relate those statistics to our monthly observations of capacity in ERCOT. Strategy 3 uses the 15- and 5-minute market data to look for real-time trader responses to the Incentive. Additionally, we use daily economic data from the Bureau of Labor and Statistics and weather data for Texas from NOAA to calculate monthly economic and climatic controls and annual Levelized Cost of Electricity data from Berkeley National Laboratory to control for renewable installation costs.

Table \ref{TBL_sum} describes the mean values of the ERCOT real-time market data, including select values of capacity from EIA Form 860M and weather from NOAA. Statistics are grouped into three Incentive ranges: (1) all intervals, (2) intervals where the Incentive is inactive, and (3) those where it is active. Producers receive the Incentive in addition to the Energy Only Price.

The average Incentive was \$11.00, \$0.00, and \$55.31 during All, Inactive, and Active intervals, respectively. The Incentive is positive (Incentive Active) in 20\% of all intervals. The energy price averaged \$30.70, decreasing to \$21.84 during inactive intervals and increasing to \$66.38 during active intervals.

\begin{table}[!htbp]\centering
\def\sym#1{\ifmmode^{#1}\else\(^{#1}\)\fi}
\caption{Means of ERCOT Real-time Market by Incentive State}
\small
\begin{tabular}{l*{3}{r}}
\hline\hline
\\[-1.8ex] 
                     &\multicolumn{1}{c}{All Intervals}&\multicolumn{1}{c}{Incentive Inactive}&\multicolumn{1}{c}{Incentive Active}\\
                    &\multicolumn{1}{c}{(1)}&\multicolumn{1}{c}{(2)}&\multicolumn{1}{c}{(3)}\\
                   
\hline
\hline
\\[-1.8ex]
Price - \$/MW               &       30.70&       21.84&       66.38\\
Incentive Active - share              &        0.20&        0.00&        1.00\\
Incentive - \$/MW             &       11.00&        0.00&       55.31\\
Supply Nat. Gas - GW   &        4.76&        4.29&        6.66\\
Supply Renewable - GW &        2.38&        2.52&        1.85\\
Supply Other - GW &        3.70&        3.59&        4.15\\
Capacity Utilization - \%     &       43.30&       40.13&       56.08\\
Reserves - GW         &       14.61&       15.75&       10.03\\
Nat. Gas Capacity - GW                &       56.47&       56.45&       56.57\\
Renewable Capacity - GW        &       26.78&       26.48&       28.01\\
Other Capacity - GW            &       22.12&       22.16&       21.98\\
Cent. Midpoint Temp. - $^\circ C$  &       -0.00&        0.18&       -0.72\\
Cent. Midpoint Temp. Sq. - $^\circ C$    &       43.54&       41.98&       49.81\\
Nat. Gas Price - \$/MMBtu             &        3.32&        3.21&        3.74\\
Wind speed - Km/hr       &        0.88&        0.99&        0.45\\
Year                &     2018.66&     2018.59&     2018.92\\
Month               &        6.61&        6.60&        6.67\\
Hour                &       11.50&       10.75&       14.52\\
Minute              &       22.50&       22.41&       22.84\\
\hline
Observations        &      257,252&      206,096&       51,156\\
\hline\hline
\end{tabular}
\label{TBL_sum}
\end{table}

Natural gas provided an average of 4.76 GW of supply during all intervals, 44\% of total supply, while renewables provided 2.3 GW of supply, 22\% of the total. During active intervals, supply by natural gas generators increased by 1.90 GW, and supply by renewable generators decreased by 0.53 GW on average. The average Capacity Utilization, the ratio of demand over demand plus remaining reserves, was 43\%, 40\%, and 56\% for all inactive and active intervals, respectively. Reserves averaged 14.61 GW across all intervals and 15.75 GW and 10.03 GW during inactive and active, respectively. Capacity averaged 56.47 GW, 26.78 GW, and 22.12 GW across all intervals for natural gas, renewables, and other generating sources, respectively, varying only slightly in inactive and active intervals. The average midpoint temperature was $0.18^\circ C$ above the mean during inactive intervals and $-.72^\circ C$ below the mean for active intervals. The natural gas price increased from \$3.21 during inactive intervals to \$3.74 during active intervals. Average wind speeds decrease from 0.99 km/hr during inactive intervals to 0.45 km/hr in active intervals. Finally, active intervals are approximately 4 hours later in the day on average than inactive intervals, as the hour of active intervals averages 2:30 pm while inactive intervals average approximately 10:45 am.

Using our data, we will examine whether the Incentive increased natural gas capacity in ERCOT and how traders responded in real-time when the Incentive was active. However, before exploring these questions empirically, we check the policy's theoretical impact on a profit-maximizing firm's decision to enter or exit the market.

\section{Strategy 1: Theoretical Impact on Market Entry}

Figure \ref{fig:enter-label2} suggests some market entry has occurred since the beginning of the Incentive program, but not from dispatchable NG generators initially targeted by the program and, therefore, not at the critical sunset hour demand. To explore whether or not the program's design encourages market entry, we use a classic partial equilibrium framework and compare the firm's profit-maximizing output with and without the Incentive. The Incentive increases as energy reserves become scarce, that is, as energy generation (demand) grows relative to total generating capacity (supply). We will consider both the monopoly and competitive market cases. 

\begin{center}
\hspace*{-3cm}\begin{tikzpicture}

\begin{axis}[
scale = 0.8,
xmin = 0, xmax = 11,
ymin = 0, ymax = 11,
axis lines* = left,
xtick = {0}, ytick = \empty,
axis on top,
clip = false,
]

\fill[Gray, opacity = 0.4] (0, 7) -- (5, 7) -- ( 7, 5.5) -- (0, 5.5);
\fill[LimeGreen, opacity = 0.2] (0, 8.5) -- (7, 8.5) -- ( 5, 7) -- (0, 7);

\addplot[color = Gray, dashed] coordinates {(0, 8.5) (7, 8.5)};
\addplot[color = Gray, dashed] coordinates {(0, 7) (5, 7)};
\addplot[color = Gray, dashed] coordinates {(0, 5.5) (7, 5.5)};
\addplot[color = Gray, dashed] coordinates {(5, 0) (5, 7)};
\addplot[color = Gray, dashed] coordinates {(7, 0) (7, 5.55)};

\addplot[color = black, very thick] coordinates {(1,10) (9,4)};
\addplot[color = LimeGreen, very thick] coordinates{(1,4) (9,10)};
\addplot[color = LimeGreen, opacity = 0.45, very thick] coordinates{(1,1) (9,7)};


\node [right] at (current axis.right of origin) {$Q$};
\node [above] at (current axis.above origin) {$P$};
\node [below right=0pt and 5pt] at (6.8, 7.9) {$S$};
\node [right] at (9.1, 4.2) {$P$};
\node [right] at (9.1, 10.4) {$MC$};
\node [right] at (9.1, 7.4) {$MC-S$};
\node [left] at (0, 8.6) {$P_P$};
\node [left] at (0, 7) {$P_E$};
\node [left] at (0, 5.4) {$P_C$};
\node [below] at (5.2, 0) {$Q_E$};
\node [below] at (7.2, 0) {$Q_S$};
\node [right, align=center] at (3, 12) {Firm subsidy $S$};
\node [above, align=center] at (6.25, -3.9) {(a)};

\draw[-{Triangle[length=4mm, width=2mm]}, red, opacity = 0.4] (7, 7.8) to (7, 6);
\draw[-{Triangle[length=4mm, width=2mm]}, red, opacity = 0.4] (5.5, -1.75) to (7, -1.75);

\end{axis}

\begin{axis}[
scale = 0.8,
xmin = 0, xmax = 11,
ymin = 0, ymax = 11,
axis lines* = left,
xtick = {0}, ytick = \empty,
axis on top,
clip = false,
shift = {(axis cs: 16, 0)},
]

\fill[Gray, opacity = 0.4] (0, 7) -- (0, 5.5) -- (7, 5.5) -- (5, 7);
\fill[LimeGreen, opacity = 0.2] (0, 5.5) -- (0, 4) -- (5, 4) -- (7, 5.5);

\addplot[color = gray, dashed] coordinates {(0, 7) (5, 7)};
\addplot[color = gray, dashed] coordinates {(0, 5.5) (7, 5.5)};
\addplot[color = gray, dashed] coordinates {(0, 4) (5, 4)};
\addplot[color = gray, dashed] coordinates {(5, 0) (5, 7)};
\addplot[color = gray, dashed] coordinates {(7, 0) (7, 5.5)};

\addplot[color = black, very thick] coordinates {(1,10) (9,4)};
\addplot[color = black, opacity = 0.3, very thick] coordinates {(1,7) (9,1)};
\addplot[color = LimeGreen, very thick] coordinates{(1,1) (9,7)};


\node [right] at (current axis.right of origin) {$Q$};
\node [above] at (current axis.above origin) {$P$};
\node [below right=0pt and 5pt] at (7.8, 3.7) {$T$};
\node [right] at (9.1, 4.2) {$P$};
\node [right] at (9.1, 1.3) {$P-T$};
\node [right] at (9.1, 7.4) {$MC$};
\node [left] at (0, 7) {$P_C$};
\node [left] at (0, 5.4) {$P_E$};
\node [left] at (0, 4) {$P_P$};
\node [below] at (5.2, 0) {$Q_T$};
\node [below] at (7.2, 0) {$Q_E$};
\node [right, align=center] at (3, 12) {Consumer tax $T$};
\node [above, align=center] at (6.25, -3.9) {(b)};

\draw[-{Triangle[length=4mm, width=2mm]}, red, opacity = 0.4] (8, 4.0) to (8, 2.2);
\draw[-{Triangle[length=4mm, width=2mm]}, red, opacity = 0.4] (7, -1.75) to (5.5, -1.75);

\end{axis}

\end{tikzpicture}\hspace*{-3cm}
\end{center}
\textbf{Figure 4:} Impact on competitive partial equilibrium output $Q$ from a firm subsidy $S$, panel (a), and consumer tax $T$, panel (b).
\newline


Figure 4(a) shows how a unit subsidy $S$ shifts the supply curve $MC$ down by the subsidy amount, increasing the competitive equilibrium quantity exchanged from $Q_E$ to $Q_S$ and raising the producer price received to $P_P$. The consumer pays a price $P_C$, which is lower than the original equilibrium price $P_E$ but by less than the full subsidy amount. Consumer and producer welfare gains are shaded gray and green, respectively.

Panel (b) shows how a unit tax $T$ shifts the aggregate demand curve lower by the tax amount, lowering equilibrium output to $Q_T$ and raising the total cost to the consumer to $P_C$. The producer receives price $P_P$, lower than the original equilibrium price $P_E$ but by less than the total tax amount. Consumer and producer welfare losses are shaded gray and green, respectively.

It appears the quantity reduction under the tax may fully offset the expansion under the subsidy. We will formalize this result for the competitive and monopoly markets in the Lemmas subsection after a few preliminaries. 

\subsection{Preliminaries} 

Departing from the commonly studied subsidy scenarios \citep{Jenkin}, the consumer pays the ERCOT Incentive directly to the producer, effectively creating an energy tax $T$ on the consumer and subsidy $S$ to the producer, the two amounts equal $S=T$.

In the ERCOT real-time energy market, consumers and producers submit bid and offer schedules, listing prices and quantities, at the beginning of each 15-minute trading period. Clearing prices are calculated, and contracts are settled at the end of each period. The level of energy scarcity is measured every five minutes and the Incentive is calculated based on the average of the last three five-minute scarcity values. In this analysis, we will ignore the time subscript on the subsidy amount $S$ for two reasons: First, ERCOT's 15-minute ahead forecast is intended to be accessible to all market participants and accurate enough to make short-term discrepancies between the forecasted and actual payments uninteresting; Second, we will show our equilibrium result does not depend on the specification of the Incentive, so long as the tax and subsidy amounts remain identical.

In the standard monopoly case, buyers negotiate to purchase quantity $Q$ at a price $P(Q)$ from a profit-maximizing firm with a cost of production $C(Q)$. Under the standard partial equilibrium assumptions, the profit $\pi$ for the firm can be formulated as total revenue less total cost
\begin{align} \nonumber
      \pi = P(Q)Q - C(Q), 
\end{align}  
with an equilibrium price solution $P_E$ equal to the difference between marginal cost and marginal revenue,
\begin{align} \label{eqn:PE}
      P_E = C'(Q) - P'(Q)Q. 
\end{align}  

The monopoly price is above marginal cost in proportion to the concavity of consumer preferences, observable in the balancing equation,
\begin{align} \nonumber
      P_E = C'(Q)\frac{1}{1+\eta},
\end{align}  
where $\eta$ is the inverted own-price elasticity $1/\epsilon$. 
Equilibrium output can be written in closed form:
\begin{align} \label{eqn:QE}
      Q_E = \frac{C'(Q) - P(Q)}{P'(Q)}.
\end{align}

In the perfectly competitive case, in contrast, equilibrium price $P_E$ cannot be written as a closed form function of $Q$, but equilibrium quantity satisfies the condition that price equals marginal cost:
\begin{align} \nonumber
      P_E = C'(Q).
\end{align}

To examine a subsidy $S$ paid by an external source to the monopoly firm for each unit of output, the profit function includes revenues from both the consumer and the subsidy: 
\begin{align}  \label{eqn:Spi}
      \pi = P(Q)Q + SQ - C(Q).
\end{align}  
The equilibrium price under the subsidy $P_S$, paid by the consumer, is now the difference between marginal cost and marginal revenue, less the subsidy,
\begin{align} \nonumber
      P_S = C'(Q) - P'(Q)Q - S. 
\end{align}  
As shown in Figure 4(a), the price paid by the consumer is lower than the non-subsidized case, in equation (\ref{eqn:PE}), but by a smaller amount than the full subsidy $S$, in proportion to the price elasticity. This can be seen from the balancing equation,
\begin{align} \nonumber
      P_S = C'(Q)\frac{1}{1+\eta} - S\frac{1}{1+\eta}.
\end{align}  
Equilibrium output can be expressed in terms of the subsidy,
\begin{align} \label{eqn:QS}
      Q_S = \frac{C'(Q) - P(Q)}{P'(Q)}+\frac{S}{P'(Q)},
\end{align}  
which is greater than the non-subsidized monopoly quantity in equation (\ref{eqn:QE}) for a positive subsidy in proportion to the concavity of consumer preference.

In the perfectly competitive case, no closed-form solution exists for equilibrium quantity. Profit is maximized at the quantity that equates the consumer paid price with marginal cost, less the subsidy:
\begin{align} \nonumber
      P_S = C'(Q) - S.
\end{align}  
This price is lower than the non-subsidized competitive equilibrium price by the full subsidy $S$, suggesting the subsidy pass-through to the consumer is 100\% (\citeauthor{Weyl2013}, 2013). 

A tax $T$ levied against the monopoly firm by an external source acts as a negative subsidy, replacing the subsidy amount $S$ with a negative tax amount $-T$. Now, the firm pays the production cost and tax: 
\begin{align} \nonumber
      \pi = P(Q)Q - C(Q) - TQ.
\end{align}  
The total equilibrium price facing the consumer is higher under the tax 
\begin{align} \nonumber
      P_T = C'(Q) + T - P'(Q)Q, 
\end{align} 
and quantity is lower
\begin{align} \nonumber
      Q_T = \frac{C'(Q) - P(Q)}{P'(Q)}-\frac{T}{P'(Q)}.
\end{align} 

Fortunately, equilibrium prices $P$ and outputs $Q$ do not depend on which party physically receives or pays the subsidy or tax. This is the Principle of Physical Neutrality first examined in competitive markets by \citeauthor{Jenkin} (\citeyear{Jenkin}) and extended to imperfectly competitive markets by \citeauthor{Weyl2013} (2013). This principle can be demonstrated by comparing the profit functions under the firm subsidy and consumer subsidy cases. 

When the consumer receives a subsidy, the monopolistic firm first observes that consumer demand has been shifted upward by the subsidy, from $P(Q)$ to $(P(Q)+SQ)$, opposite to the consumer tax case in Figure 4(b). With this in mind, the profit function is formed:
\begin{align} \nonumber
      \pi = P(Q)Q + SQ - C(Q).
\end{align}  
This matches exactly the objective function in the firm-subsidy model of equation (\ref{eqn:Spi}), and the subsidy output solution of equation (\ref{eqn:QS}) follows directly. The principal holds, similarly, for subsidy and taxation in both the monopoly and competitive cases.

\subsection{The cancelled subsidy lemmas} 
We may now turn to the paper's central question and ask if the Incentive, paid by consumers to producers, leads to market entry, specifically, should equilibrium energy capacity increase under the Incentive?

Define the equilibrium price and quantity solution under a simultaneous tax and subsidy, $P_{ST}$ and $Q_{ST}$, respectively, and the equilibrium solution pair under no subsidy or tax $P_{E}$ and $Q_{E}$. Under the standard equilibrium assumptions, the following lemma holds.\\
\hfill\\
\textit{\textbf{Lemma 1}}: The monopolistic partial equilibrium solution is unchanged by a fixed, identical, unit subsidy $S$ and tax $T$, regardless of the physically responsible parties: $Q_{ST} = Q_E \:\: \text{and } \:P_{ST} = P_E$.\\

To check this condition, we can form the profit function for a monopoly firm receiving a subsidy $S$, which is paid directly and instantaneously by a consumer tax $T$:
\begin{align} \nonumber
      \pi = P(Q)Q + SQ - TQ - C(Q).
\end{align} 
Here, the firm observes that consumer demand has shifted downward by the total tax $TQ$ and that firm revenues have increased by the subsidy payment $SQ$.

Because the payment is made directly and instantaneously between the two parties, the tax and subsidy are identical $T=S$, and the two products cancel. The profit function is simplified to
\begin{align}  \nonumber
      \pi = P(Q)Q - C(Q),
\end{align} 
an objective function identical to the standard monopoly problem, with an identical equilibrium solution pair, $P_E$ and $Q_E$ from equations (\ref{eqn:PE}) and (\ref{eqn:QE}), respectively, demonstrating our result. 

The outcome is independent of which entity physically bears the subsidy or tax, according to the Principle of Physical Neutrality, since the positive and negative payments are canceled if held by any single party or if the parties are reversed. 

The result also holds for the firm operating in a competitive market, which we state next.\\
\hfill\\
\textit{\textbf{Lemma 2}}: The competitive partial equilibrium solution is unchanged by a fixed, identical, unit subsidy $S$ and tax $T$: $Q_{ST} = Q_E \:\: \text{and } \:P_{ST} = P_E$.\\

The only change for the competitive case is a fixed price $P$ specified in the profit function: 
\begin{align} \nonumber
      \pi = PQ + SQ - TQ - C(Q).
\end{align} 

Finally, the above results hold for any functional form of tax and subsidy, including uncertain or lagged payments, since the two payments are canceled under identical specifications. To formalize this, consider subsidy and tax functions $S(X_t)$ and $T(X_t)$, respectively, where $X_t$ is a random variable with the usual triplet ($\Omega, P,\mathscr{F}_t$), with $\Omega$ the set of all possible events, $P$ the probability rule, and $\mathscr{F}_t$ the filtration containing all information up to time $t$. The following then holds.\\

\hfill\\
\textit{\textbf{Lemma 3}}: The partial equilibrium solution is unchanged by any identical subsidy and tax functions $S(X_t)$ and $T(X_t)$, respectively: $Q_{ST} = Q_E \:\: \text{and } \:P_{ST} = P_E$.\\

To verify, we can reform the profit function for the monopoly firm with subsidy function $S(X_t)$, paid for by consumer tax function $T(X_t)$:
\begin{align} \nonumber
      \pi = P(Q)Q + S(X_t) - T(X_t) - C(Q),
\end{align} 
which cancel, so long as $S(X_t)=T(X_t)$. Independence from the physically responsible party follows from \textit{Lemma 1} and extends to the competitive market from \textit{Lemma 2}.

\section{Strategy 2: Empirical Impact on Market Entry}

The third canceled subsidy lemma suggests the Incentive should not impact entry into the ERCOT generation market. To test this empirically, we estimate the Incentive's direct and total effects using a structural model of the firm's decision to enter ERCOT's applicant pool and eventually proceed into production.

\subsection{Model}

Before firms introduce generating capacity $G_{t}$ to the market in time $t$, they must first enter the ERCOT applicant pool $A_{t-n}$ some $n$ months prior. Firms apply after assessing market and climate conditions $X_{t-n}$ and the Incentive $I_{t-n}$. Twelve months ($n=12$) is within the time range firms spend in the pool before either withdrawing from the pool or entering production.\footnote{Many other lags were tested for robustness with little impact on conclusions; see Variable Lags Appendix for more detail.} The Incentive, observed by prospective firms, is also determined by market and climate conditions $X_{t-n}$. These three steps in the structural model are then:
\begin{align} 
   I_{t-12} &= \alpha_{X, I} X_{t-12} +  \pmb{\upsilon}_{I, \,t-12} \\
   A_{t-12} &= \alpha_{X, A} X_{t-12} + \alpha_{I, A} I_{t-12} + \pmb{\upsilon}_{A,\,t-12} \label{eqn:OLSA}\\
   G_{t} &= \alpha_{X, G} X_{t-12} + \alpha_{I, G} I_{t-12} + \alpha_{A, G}  A_{t-12} + \pmb{\upsilon}_{G_{t}} \label{eqn:struct}.
\end{align} 

Coefficients are estimated consistently with Ordinary Least Squares (OLS), and the parameters are interpreted as the usual direct effects. Our coefficient of interest is the market entry coefficient $\alpha_{G, I}$, representing the increase in megawatts of generating capacity $G_{t}$ at time $t$ for each dollar of the Incentive $I_{t-12}$ paid out during period $t-12$, holding all other variables constant. A positive and significant entry coefficient estimate would reject the no-entry hypothesis posed by the canceled subsidy Lemma 3.

Because we expect the applicant pool $A$ to respond causally to the incentive $I$ and market controls $X$, as in equation (\ref{eqn:OLSA}), we would expect multicollinearity to reduce statistical significance. In addition, coefficient signs would be less reliable since coefficient covariance is negative when controls are correlated. It turns out we can eliminate multicollinearity and its statistical impacts when we obtain the reduced form.

Solving the system of equations in terms of orthogonal residuals, we find the reduced form:
\begin{align} 
       I_{t-12} &= \beta_{X, I} X_{t-12} + \pmb{\upsilon}_{I, \,t-12}
       \label{eqn:RFI}\\
       A_{t-12} &= \beta_{X, A} X_{t-12} + \beta_{I, A} \, \pmb{\upsilon}_{I, \,t-12} + \pmb{\upsilon}_{A,\,t-12} \label{eqn:RFA}\\
       G_{t} &= \beta_{X, G} X_{i,t-12} + \beta_{I, G} \, \pmb{\upsilon}_{I, \,t-12} + \beta_{A, G} \, \pmb{\upsilon}_{A,\,t-12} + \pmb{\upsilon}_{G_{i,t}}.\label{eqn:RFG}
    \end{align}  

Coefficients now represent the total effect of each independent variable on the dependent variable. In our case, the entry coefficient $\beta_{I, G}$ is the total effect on generating capacity from each Incentive dollar:
\begin{align}  
       \beta_{I, G} &=  \alpha_{I, G} + \alpha_{I, A} \, \alpha_{A, G} \label{eqn:Lemma}.
    \end{align}  
The total effect of the Incentive consists of its direct effect on generating capacity $\alpha_{I, G}$, plus its impact on the applicant pool $\alpha_{I, A}$, moderated by the propensity of applicants to move into production $\alpha_{A, G}$. A positive and significant total effect estimate of $\beta_{I, G}$ would also reject the no-entry hypothesis of the canceled subsidy Lemma 3. However, this would be a weaker indication since rejection can be due to the Incentive's direct effect on market entry, its indirect impact on the applicant pool, or both.

Reduced-form coefficients can be estimated consistently with OLS equation-by-equation, recursively replacing each of the three variables with the corresponding residual as shown by \citeauthor{Wold1960} (\citeyear{Wold1960}) and \citeauthor{Pearl2000} (\citeyear{Pearl2000}). By exploiting the temporally separated and recursive nature of the structural system and noticing that the substitution of residuals in the reduced-form system corresponds to the Gram-Schmidt orthogonalization process, we can go one step beyond consistency to show that our covariate estimate of $\beta_{I, G}$ in equation (\ref{eqn:Lemma}) is an unbiased estimate of the total effect $\alpha_{I, G} + \alpha_{I, A} \, \alpha_{A, G}$, free from multicollinearity, and more efficient than the OLS estimate. 

Assume $X \subseteq \mathcal{R}^{[N \times K]}$ is a full-rank, ordered regressor matrix including dependent variable $G_t = x_K$; $X$ has positive degrees of freedom $N>K$; the error vectors have zero conditional means, $E[\upsilon_i|x_j] = 0$, $j=1,...,i-1$, and are free of autocorrelation and heteroscedasticity, $E[\upsilon_i\upsilon_i'|\mathbf{X}]=\sigma_i^2\mathbf{I}$. Referring to the proposed estimation model as Gram-Schmidt least squares (GSLS), define the GSLS coefficient estimates $c_{i,j}$, OLS coefficient estimates $b_{i,j}$, and OLS residuals $u_i$.

\hfill\\
\textit{\textbf{Theorem 1}}:
    Under the above assumptions, GSLS estimation of the reduced-form system in equations (\ref{eqn:RFI}) - (\ref{eqn:RFG}) achieves the following properties: 

    \begin{tabular}{l l l}
        (A) & Regressors are orthogonal, $u_i \perp u_j$; \\
        (B)  & Coefficients are stable,  $Cov[c_{ij},c_{ji}] = 0$;  \\
        (C)  & Coefficients are unbiased,  $E[c_{ij}]=\beta_{ij}$; and \\
        (D)  & GSLS is more efficient than OLS,  $V[c_{ij}] \leq V[b_{ij}]$. 
   \end{tabular}

\hfill\\ A proof is provided in the appendix. 

To avoid endogeneity bias, whenever covariates are determined simultaneously, such as temperature and wind speed or online and offline capacity, these covariate groups are regressed upon all prior covariates but not upon one another. This way, across-group multicollinearity will be removed, and only within-group multicollinearity will remain. In the following sections, we will apply this model to all ordered datasets and present total effects from the GSLS model alongside direct effect estimates from OLS. 

\subsection{Results and discussion - Market Entry}

Columns 1 and 2 of Table \ref{TBL_GEN} show direct and total effects for a select set of covariates on natural gas generation capacity operating in the market. Neither the direct nor total Incentive effects $\alpha_{I, G}$ and $\beta_{I, G}$, estimated by the Incentive coefficients, are significant, failing to reject the canceled subsidy hypothesis of Lemma 3. The only highly significant direct effect on natural gas capacity is wind speed, where a one km/hr increase in the 3-month rolling mean (lagged 12 months) causes 2.4 GW of capacity to exit the market, suggesting competition between wind and natural gas generation. Wind speed's total effect on the exit of natural gas capacity is lower than the direct effect, tempered by downstream economic and market influences, yet still causes 0.92 GW to exit for each additional average kilometer-per-hour at 99.9\% confidence. Increases in Offline Capacity have a positive and somewhat significant effect, both directly (95\% confidence) and in total (99\% confidence). In further examining total effects, inflation reduces natural gas capacity by 115 MW per inflation percentage point when accounting for downstream impacts, and the Shadow Price of bringing online one more offline generating unit increases capacity by 32 MW per \$1 increase in marginal generation cost. Lastly, there is a slightly significant (95\% confidence) positive total effect from increased Capacity Utilization, suggesting that scarcity, as measured through sustained decreases in reserves, may encourage the entry of natural gas capacity in total, if not directly.

\begin{table} 
    \centering
    \footnotesize
    \captionsetup{skip=0pt} 
    \captionsetup{belowskip=0pt} 
    \caption{Total Megawatts Generating - Select Covariates} 
    \setlength{\tabcolsep}{0pt}
    \begin{tabular}{@{\extracolsep{0pt}}lD{.}{.}{-1} D{.}{.}{-1} D{.}{.}{-1} D{.}{.}{-1} } 
        \\[-1.8ex]\hline 
        \hline \\[-1.8ex]
        \multicolumn{2}{c}{}\\
        Dep. Var.: \\ Generating Cap. (MW) & \multicolumn{2}{c}{Natural Gas} & \multicolumn{2}{c}{Renewables} \\
        \cmidrule(lr){2-3}\cmidrule(lr){4-5}
        & \multicolumn{1}{c}{Direct} & \multicolumn{1}{c}{Total} & \multicolumn{1}{c}{Direct} & \multicolumn{1}{c}{Total}\\ 
        \\[-1.8ex] & \multicolumn{1}{c}{(1)} & \multicolumn{1}{c}{(2)} & \multicolumn{1}{c}{(3)} & \multicolumn{1}{c}{(4)}\\ 
        \hline \\[-1.8ex] 
Midpoint Temp. - $^\circ$C 12mo mean & -202.53 & -81.44 & 77.33 & -236.32^{*} \\ 
  Midpoint Temp. Sq. - $^\circ$C  12mo Mean & 182.44 & 155.23^{*} & 171.56 & 324.48^{***} \\ 
  Mean Windspeed - km/hr 12mo Mean & -2,432.79^{***} & -919.20^{***} & -956.74^{*} & -743.13^{***} \\ 
  Labor Force Pop. - Millions & 363.92 & 308.88 & 99.95 & 6,591.84^{***} \\ 
  Unemp. Rate & -47.92 & -5.49 & -22.58 & 211.92^{**} \\ 
  CPI & -82.50 & -115.02^{**} & -30.12 & 50.35 \\ 
  Interest Rate & 369.12 & 368.29 & 238.81 & -1,636.90^{***} \\ 
  Cost Wind Installation - \$/MWh & -64.63 & -42.05 & 51.45 & -264.90^{***} \\ 
  Cost PV Installation - \$/MWh & 2.22 & 37.54 & -26.43 & 136.85^{***} \\ 
  Natural Gas Price - \$/MMBtu & -436.63 & -17.21 & -322.33 & 112.17 \\ 
  Online Capacity - GW & 10.44 & 37.51 & 103.30 & -301.11^{***} \\ 
  Offline Capacity - GW & 919.01^{*} & 804.84^{**} & -617.20 & -665.78^{*} \\ 
  Capacity Utilization - \% & 475.85 & 772.47^{*} & -1,162.50^{*} & -1,023.50^{**} \\ 
  Shadow Price - \$/MW & 4.45 & 32.20^{*} & -113.66 & -30.74^{**} \\ 
  Energy Only Price - \$/MW & 12.15 & 3.78 & 65.68 & 29.44^{***} \\ 
  Incentive - \$/MW & -34.13 & -21.80 & -51.55^{*} & -24.03 \\ 
  Change Peaker Net Margin - \$/GW & 1,080.57 & 1,213.13 & 884.36 & -1,326.20 \\ 
  NG GINI Index & -72.67 & -8.10 & -191.19 & 1.70 \\ 
  Wind GINI Index & 109.22 & 165.13 & 231.34 & 302.94^{*} \\ 
  Renewable Appl. Pool - MW & -0.08 & -0.09 & 0.35^{*} & 0.30^{*} \\ 
  NG Appl. Pool - MW  & -0.18 & -0.16 & -0.59^{***} & -0.69^{***} \\ 
        \hline \\[-1.8ex] 
        \hline \\[-1.8ex] 
        \textit{Note:}  & \multicolumn{4}{r}{$^{*}$p$<$0.05; $^{**}$p$<$0.01; $^{***}$p$<$0.001} \\ 
    \end{tabular} 
    \label{TBL_GEN}
\end{table} 

Columns 3 and 4 of Table \ref{TBL_GEN} display the corresponding model output for renewable generating capacity for comparison. There are two notable results in comparison to the natural gas segment. First, the Incentive continues to have a negative and insignificant or weakly significant effect on renewable capacity entry, supporting the Cancelled Incentive Lemma. Second, many more coefficients are statistically significant for renewable capacity, suggesting entry and exit may be less restricted in the renewable market in response to changing economic and climatic conditions. In total, increased inflation and installation costs are significant drivers of exit in the renewables market, while greater population and energy prices appear to encourage entry. Directly, the NG Applicant Pool coefficient is also negative and significant, suggesting some crowding out may occur between natural gas and renewables capacity. 

These results suggest that the Incentive and scarcity, more broadly, do not appear to promote natural gas capacity growth directly. The lack of responsiveness from the natural gas segment and potential crowding out of generating renewable capacity motivates our next section, where we will further examine the effects of the incentive on the applicant pools.

\subsection{Strategy 2. B: Empirical Impact on the Applicant Pool}

Past research has noted the concentration of generating capacity in ERCOT \citep{hortacsu, coneEstimates} and questioned the market's competitiveness \citep{dyer2011}, suggesting the lack of a Capacity Market in ERCOT results in higher price volatility and less reliability than long-term capacity incentives \citep{BAJOBUENESTADO2017272}. So, we explore whether investors enter the applicant pool but fail to reach generating status due to factors beyond the production costs, returns, and incentives captured by our model. Applicants representing eight GW of natural gas capacity, or about 48\% of the applicant pool, exited before reaching operation. In comparison, only three GW, or 5\%, of the renewable pool exited over the same period. However, these figures don't explain why a generator in the applicant pool failed to become operational.

Columns 1 and 2 of Table \ref{TBL_pool} compare natural gas application pool total effects from equation (\ref{eqn:RFA}) with the generating capacity total effects, already shown in Column 2 of Table \ref{TBL_GEN}. First, the applicant pool appears much more responsive to drivers of entry and exit. Wind generation continues to compete with natural gas in the applicant pool, as a one-dollar decrease in the cost of wind installation results in 0.2 GW of natural gas entry into the pool. Entry into the natural gas pool also increases with natural gas concentration (NG GINI Index), at 99.9\% confidence, yet fails to become operational (Column 2), as we saw in Table \ref{TBL_GEN}. Finally, the Incentive coefficient $\beta_{I, A}$ is insignificant consistent with the canceled incentive Lemma 3. 

\begin{table}[!h]
    \centering
    \footnotesize
    \captionsetup{skip=0pt} 
    \captionsetup{belowskip=0pt}  
    \caption{Comparison Applicant Pool to Generating - Select Covariates} 
    \setlength{\tabcolsep}{0pt}
    \begin{tabular}{@{\extracolsep{1pt}}lD{.}{.}{-3} D{.}{.}{-3} D{.}{.}{-3} D{.}{.}{-3} } 
        \\[-1.8ex]\hline 
        \hline \\[-1.8ex]
        \multicolumn{2}{c}{}\\
        & \multicolumn{2}{c}{Natural Gas} & \multicolumn{2}{c}{Renewables} \\
        \cmidrule(lr){2-3}\cmidrule(lr){4-5}
        Dependent Variable: (MW) & \multicolumn{1}{c}{Appl. Pool} & \multicolumn{1}{c}{Gen. Cap.} & \multicolumn{1}{c}{Appl. Pool} & \multicolumn{1}{c}{Gen. Cap,}\\ 
        & \multicolumn{1}{c}{(1)} & \multicolumn{1}{c}{(2)} & \multicolumn{1}{c}{(3)} & \multicolumn{1}{c}{(4)}\\ 
        \hline \\[-1.8ex] 
Midpoint Temp. - $^\circ$C 12mo mean & 781.47^{***} & -81.44 & 112.23 & -236.32^{*} \\ 
  Midpoint Temp. Sq. - $^\circ$C 12mo mean & 48.68 & 155.23^{*} & 291.71^{***} & 324.48^{***} \\ 
  Mean Wind speed - km/hr 12mo mean & -1,521.35^{***} & -919.20^{***} & -1,352.60^{***} & -743.13^{***} \\ 
  Labor Force Pop. - Millions & -5,302.81^{***} & 308.88 & 2,984.89^{***} & 6,591.84^{***} \\ 
  Unemp. Rate & -252.97^{*} & -5.49 & -81.58 & 211.92^{**} \\ 
  CPI & -39.47 & -115.02^{**} & 16.67 & 50.35 \\ 
  Interest Rate & 1,650.64^{***} & 368.29 & -1,330.31^{***} & -1,636.90^{***} \\ 
  Cost Wind Installation - \$/MWh & 297.06^{**} & -42.05 & 5.38 & -264.90^{***} \\ 
  Cost PV Installation - \$/MWh & -211.04^{***} & 37.54 & -22.49 & 136.85^{***} \\ 
  Natural Gas Price - \$/MMBtu & -767.11^{***} & -17.21 & 227.93 & 112.17 \\ 
  Online Capacity - GW & 619.91^{***} & 37.51 & -63.99 & -301.11^{***} \\ 
  Offline Capacity - GW & -2,053.85^{***} & 804.84^{**} & -351.79 & -665.78^{*} \\ 
  Capacity Utilization - \% & 368.73 & 772.47^{*} & -217.99 & -1,023.50^{**} \\ 
  Shadow Price - \$/MW & 40.73^{*} & 32.20^{*} & -33.97^{**} & -30.74^{**} \\ 
  Energy Only Price - \$/MW & -1.90 & 3.78 & 15.53^{*} & 29.44^{***} \\ 
  Incentive - \$/MW & -31.71 & -21.80 & 19.31 & -24.03 \\ 
  Change Peaker Net Margin - \$/GW & -5,397.12^{***} & 1,213.13 & -2,601.29^{**} & -1,326.20 \\ 
  NG GINI Index & 136.93^{***} & -8.10 & 26.88^{***} & 1.70 \\ 
  Wind GINI Index & -190.87 & 165.13 & -72.43 & 302.94^{*} \\ 
  Renewable Appl. Pool - MW &  & -0.09 &  & 0.30^{*} \\ 
  NG Appl. Pool - MW  &  & -0.16 &  & -0.69^{***} \\ 
        \hline 
        \hline \\[-1.8ex] 
        \textit{Note:}  & \multicolumn{4}{r}{$^{*}$p$<$0.05; $^{**}$p$<$0.01; $^{***}$p$<$0.001} \\ 
    \end{tabular} 
    \label{TBL_pool}
\end{table} 

To validate these results, Columns 3 and 4 of Table \ref{TBL_pool} make the same comparison between the applicant pool capacity and generating capacity in the renewables segment. This comparison reinforces the insignificance of the renewable segment to the Incentive. Further, the applicant pool and generating capacity respond negatively and significantly to the Shadow Price, contracting by 36 MW and 31 MW, respectively, for each dollar increase in marginal cost. Growth in the applicant pool is positive and significant as the concentration of natural gas resources increases (Column 3). However, similar to natural gas, this growth is not realized as operational capacity (Column 4).

The results from our comparison of the applicant pool reinforce findings from Strategy 2, namely, the Incentive is not a significant cause of entry.

\section{Strategy 3: Market Participant Response}

The lack of natural gas entry in response to the Incentive in Strategy 2a and the negative response in 2b suggest the Incentive fails to attract and may be discouraging capacity growth. The canceled subsidy Lemma 3 suggests market participants will optimally lower their bids and offers to offset the Incentive. But how is this accomplished in practice? ERCOT adds the Incentive to the energy price at the end of each 15-minute trading period. A perfect offset would require market participants to have excellent information about the Incentive and to be able to adjust bids and asks incrementally lower, all in near real-time. In this section, we empirically test for evidence of a bid or ask response to the Incentive.

\subsection{Model}

To test whether market participants actively lower their bids, either directly or indirectly, in response to the Incentive, we will allow for either a fixed response $\beta_1$, an amount by which participants may lower their bid whenever the Incentive is active, $Incentive_t > 0$, or a proportional response $\beta_2$, an amount participants may lower their bid for each dollar paid through the Incentive. 

To test for these two possible responses simultaneously, we estimate direct and total effects for the following model:
\begin{align} 
     \label{eqn:response}
      P_t = \beta_1 \mathbb{1} (Incentive_t > 0)+ \beta_2Incentive_t + \alpha\textbf{X}_t + \epsilon_t.
\end{align}  

\noindent Here, $P_t$ is the clearing price of energy before the Incentive in the 15-minute interval $t$, and $X$ is a vector of annual and monthly fixed effects and the covariates described in Table \ref{TBL_sum}. 

Two extreme trading responses exactly fulfill Lemma 3. A fixed-only response, in which participants lower their bids by a fixed amount $\beta_1=$ \$55.31 whenever the Incentive is active, $\mathbb{1}(Incentive_t > 0)$. This would offset the average Incentive over all active 15-minute trading intervals with no proportional response $\beta_2=0$. Such a response might be reasonable if Incentive information is delayed or costly to incorporate into future bids, but information about the average Incentive is known. Alternatively, in a proportional-only response, participants would lower their bid by one dollar for each dollar to be paid out under the Incentive $\beta_2=-1$ in the upcoming 15-minute trading interval. This would exactly offset the Incentive with no fixed response, $\beta_1=0$. This might be reasonable when the Incentive amounts are known perfectly in real-time and easily incorporated into the bid-ask schedules. A combination of these two strategies could also fully cancel out the Incentive in line with Lemma 3.

An imbalance exists between the number of active and inactive intervals in our data set, as shown in the summary statistics in Table \ref{TBL_sum}. This introduces a potential weighting error that may bias our OLS estimates \citep{Ruben1973}. Table \ref{TBL_sum} shows that the Incentive is active in only 20\% of intervals. To check for potential weighting bias, we rebalance the treatment and control set using a covariate matching method that minimizes the Mahalanobis distance between the means and variances of characteristics in the two groups \citep{DehejiaWahba2002}. We then re-estimate equation (\ref{eqn:response}) for the average (direct) treatment effect (ATET), $\beta_1$, with the re-weighted data in Table \ref{TBL_match}.

\subsection{Results and discussion}

Direct and total effects for the two responses in Equation \ref{eqn:response} are presented in Table \ref{B} for selected covariates with the original observation weighting. 

\begin{table}[!h]
  \caption{Underbidding Direct and Total Effects - Fixed Effects Omitted} 
  \label{B} 
\small
\begin{tabular}{@{\extracolsep{1pt}}lD{.}{.}{-3} D{.}{.}{-3} } 
\\[-1.8ex]\hline 
\hline \\[-1.8ex] 
 & \multicolumn{2}{c}{\textit{Dependent variable:}} \\ 
\cline{2-3} 
\\[-1.8ex] & \multicolumn{2}{c}{Energy Price} \\ 
& \multicolumn{1}{c}{Direct} & \multicolumn{1}{c}{Total}\\ 
 & \multicolumn{1}{c}{(1)} & \multicolumn{1}{c}{(2)}\\ 
\hline \\[-1.8ex] 
 Nat. Gas Capacity - GW & -11.25^{***} & -24.51^{***} \\ 
  Renewable Capacity - GW & -17.15^{***} & -24.28^{***} \\ 
  Other Capacity - GW & -17.58^{***} & 67.87^{***} \\ 
  Midpoint Temp.  & -2.29^{***} & -2.15^{***} \\ 
  Midpoint Temp. Sq. & 0.63^{***} & 0.76^{***} \\ 
  Wind speed km/hr & -1.03^{***} & -1.33^{***} \\ 
  Natural Gas Price - \$/MMBtu & 32.36^{***} & 31.01^{***} \\ 
  Supply Nat. Gas - GW & -63.35^{***} & 16.37^{***} \\ 
  Supply Renewables - GW & -57.07^{***} & 0.63 \\ 
  Supply Other - GW & -106.98^{***} & -15.20^{***} \\ 
  Reserve Capacity - GW & 36.57^{***} & -7.35^{***} \\ 
  Capacity Utilization - \% & 31.68^{***} & 23.73^{***} \\ 
  Incentive Active - Y/N & -79.97^{***} & -79.13^{***} \\ 
  Incentive Payment - \$/MW & -0.05^{***} & -0.05^{***} \\
 \hline \\[-1.8ex] 
\hline 
\hline \\[-1.8ex] 
\textit{Note:}  & \multicolumn{2}{r}{$^{*}$p$<$0.05; $^{**}$p$<$0.01; $^{***}$p$<$0.001} \\ 
\end{tabular} 
\label{TBL_Under}
\end{table} 

The results suggest market participants may use a mixed strategy of fixed and proportional bid offsets. The direct fixed offset $\beta_1$ (Incentive Active)  is a negative and significant \$79.97, with an additional proportional offset of \$0.05 for every dollar increase in the incentive. The combined direct offset of \$82.74 is \$27.43 greater than the amount required to fully offset the average Incentive of \$55.31 from Table 2. The total offset of \$79.13 is \$23.82 greater than the amount required to match the average Incentive.  The apparent overreaction suggests traders may be highly aware of the need to cancel out the pending Incentive, in line with Lemma 3, but lack accurate real-time information about the magnitude and timing of the Incentive. The \$82.74 drop is more in line with the extreme Incentive payouts, such as those seen during the 2021 Texas energy crises, than with normal scarcity conditions.\footnote{Results for periods excluding Winter Storm Uri and other outlier events are included in the appendix. }

Other dynamics in the Texas energy market appear to function normally. Additional capacity from both natural gas and renewable sources lowers the energy price significantly, both directly and in total. Increases in capacity from Natural Gas and Renewables have a similarly significant and negative effect on the energy price, with an additional GW of capacity reducing the energy price by \$11.25 and \$17.15, respectively. Wind has a significant and negative impact on the energy price, with a one km/hr increase in wind speed reducing the energy price by \$1.03 per MW. This price-depressing effect of wind speed increases  slightly when accounting for downstream covariates, such as natural gas price and the Reserve Capacity in the system, reducing the energy price by \$1.33 in total. The market is responsive to natural gas prices, increasing \$32.36 per MW for every \$1.00 increase in the price of natural gas. The total effect is somewhat depressed at \$31.01, accounting for impacts on reserve capacity and other downstream covariates. Scarcity has a positive and significant effect, increasing the energy price by \$31.68 per MW for every one percentage-point increase in generating capacity utilization.

\begin{table}[!h]
\centering
\caption{Incentive Active Effect on Energy Price With Matched Data}
\small
\begin{tabular}{l*{4}{r}}
\hline\hline \\[-1.8ex]
            &           Coefficient&       Std. error&           z&      P\\
\hline \\ [-1.8ex]
Active Incentive &   -494.744&     19.668&   -25.155&    1.2e-139\\
\\[-1.8ex]\hline 
\hline \\[-1.8ex] 
        & \multicolumn{2}{c}{Standardized Differences} & \multicolumn{2}{c}{Variance Ratio} \\
        \cmidrule(lr){2-3}\cmidrule(lr){4-5}
        & \multicolumn{1}{c}{Raw} & \multicolumn{1}{c}{Matched} & \multicolumn{1}{c}{Raw} & \multicolumn{1}{c}{Matched}\\ 
        \\[-1.8ex]\hline 
\hline \\[-1.8ex]  
Supply Nat. Gas - GW &    1.124&    0.170&    1.726&    1.176\\
Supply Renewables - GW &   -0.512&   -0.125&    0.853&    1.004\\
Supply Other - GW&    0.645&    0.095&    0.997&    1.014\\
Capacity Utilization - \% &    1.739&    0.390&    1.745&    1.248\\
Reserve Capacity - GW &   -1.592&   -0.405&    0.853&    1.232\\
Nat. Gas Capacity - GW        &    0.118&   -0.004&    1.122&    0.996\\
Renewable Capacity - GW&    0.171&   -0.007&    1.238&    0.996\\
Other Capacity - GW    &   -0.082&    0.000&    0.910&    1.002\\
Midpoint Temp. - $C^\circ$ &   -0.134&   -0.020&    1.175&     1.021\\
Natural Gas Price - \$/MMBtu    &    0.268&   -0.000&    2.058&    1.012\\
Wind speed - Km/hr &   -0.218&   -0.011&    0.909&    1.017\\
Year        &    0.148&   -0.006&    1.109&    0.993\\
Month       &     0.021&    0.002&    0.775&    0.989\\
Day         &   -.051&    0.006&    1.032&     1.022\\
Hour        &    0.609&    0.052&    .519&    0.980\\
Minute      &    0.026&     0.007&    0.984&    0.999\\
\hline\hline
\end{tabular}
\label{TBL_match}
\end{table}

To check for the potential bias from fewer active Incentive observations, Table \ref{TBL_match} shows the average treatment effect (ATET) of the fixed response when the Incentive is active (treatment) compared to inactive (control). Matched standardized differences (differences in means) are near zero except for Reserve Capacity and Capacity Utilization, which are closely tied to active Incentive conditions. Matched variance ratios are close to unity, suggesting that rebalancing reduces most differences between active and inactive Incentive probability moments despite significant differences in the original data.

Rebalanced treatment effect estimates cannot be compared directly to the OLS and GSLS results in Table \ref{TBL_Under}. By rebalancing the observations in the control and treatment groups, we also rebalance the Incentive amounts paid to the group, which may be more or less than the \$494.74 offset. Rebalancing results reinforce the idea that traders were aware of the Incentive and lowered prices by a significant amount when the Incentive was active. 

\section{Conclusion}

ERCOT's policy to ensure reliability in Texas' energy market aimed to improve reliability by increasing natural gas generating capacity by 2,000 MW. The results from our research suggest that the Incentive did not encourage entry into the Texas energy market and did not contribute to reliability. The partial equilibrium solution to the Incentive does not raise profits for generators, as intended, since market participants lower bid and ask prices and effectively cancel out the Incentive. The theoretical findings are supported by our causal analysis of generating capacity, which showed the Incentive did not contribute directly or in total to entry by natural gas generators. Finally, market participants appear to know and practice the canceling strategy, lowering bids and asking prices whenever the Incentive is active. Clearing prices were reduced by more than the actual Incentive value, suggesting traders were aware of the need to lower prices but may not have had ready access to complete Incentive information or found implementing finer price adjustments difficult or costly.

We do not examine how the program was initially designed or why it was implemented in its current form. Several alternative policies are standard in energy markets that would avoid the canceling-out effect observed in this analysis. Due to grid instability, ERCOT has faced pressure to come under greater regulatory authority from the Federal Energy Regulatory Commission. The promise of improved reliability from additional natural gas generation capacity might assuage such pressure. However, the adopted policy's unusual self-canceling feature fails to improve pricing, market competition, or grid reliability.

\newpage
\bibliographystyle{aea}
\bibliography{AER-Article}

\appendix

\section{Appendix}

This appendix provides additional data discussion, proof of Theorem 1, and robustness checks. Data and code to replicate the main results can be accessed at https://github.com/devinmounts/ercot\_incentive\_effects.

\subsection{Firm Profits}

Figure A1 shows profit margins for six ERCOT firms as reported through annual SEC filings and fiscal reports. Profit margins are calculated based on operating revenues minus expenses or earnings before interest, taxes, depreciation, and amortization (EBITDA) for Texas or ERCOT-specific generation operations. These firms account for roughly 36\% of ERCOT's natural gas capacity in 2022. While this is not a scientific analysis, comparing profit margin to annual Incentive value does not demonstrate a positive relationship.
\begin{figure}
    \includegraphics[height=6cm]{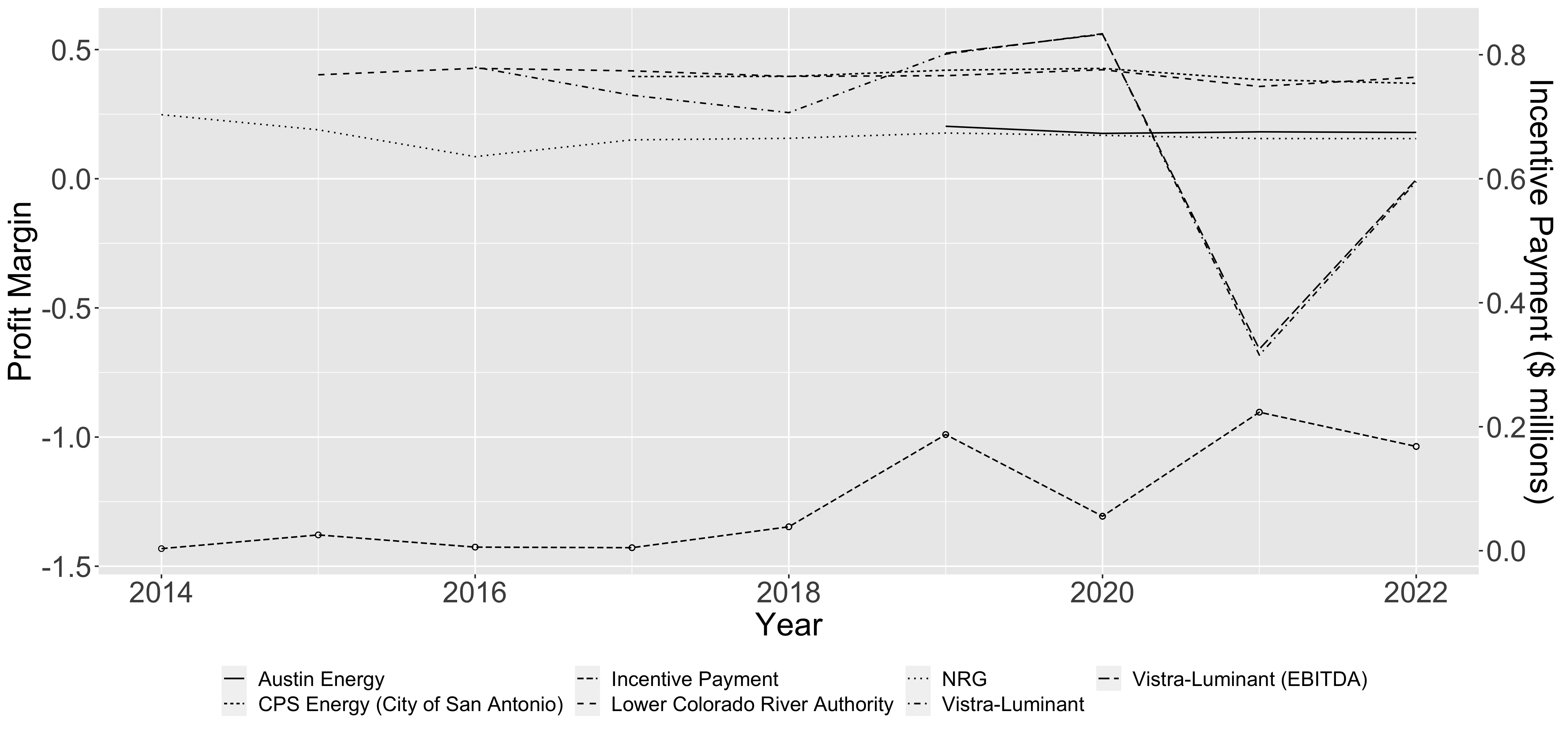}
    \caption{Firm Profits and Annual Incentive Value}
    \label{fig:enter-label3}
\end{figure}

\subsection{Sample Construction}

69\% of the 3,845 generators listed on EIA FORM 860M in Texas have null balancing authority codes at some point in the record during the analysis period. Of all null values, 92\% are concentrated in the first 12 months of the data. To identify generators within the ERCOT market, we replaced null balancing authority codes with the balancing authority identified for the same plant in future observations. Two hundred ninety-eight generators retain null balancing authority codes. 40\% of the generators with remaining null balancing authority codes are located within ERCOT's boundary. These 119 generators are assigned to ERCOT due to their geo-location. Of these 119 generators, 16, 11, and 9 operating or applicant natural gas, renewable, and other generators, respectively, come into our analysis. The sample is limited to the 2,249 generators assigned to the ERCOT balancing authority.

\subsection{Proof of Theorem 1}
The proof here is adapted from Cross and Bucolla (\citeyear{crossGSLS}).

(A) The orthogonality of the regressors is shown by \citet[pp. 57-59]{Wong}. 

(B) Stability of coefficients follows directly from their orthogonality (A) since coefficient covariance 
\begin{equation*}
    Cov[c_{ik},c_{jk}]  = \sigma_{\upsilon_k}^2 \frac{-r_{u_{i}, u_{j}} }{1-r_{u_{i}, u_{j}}^2}
\end{equation*}

\noindent is linear in regressor correlation $r_{u_i,u_j}=u_i'
u_j / \surd(u_i'u_i \; u_j'u_j)$ and zero for any recursive regressor pair.

(C) Unbiasedness. Coefficients $a_{ij}$ of reduced-form parameter matrix A can be expressed as the recursive sequence \begin{equation*}
    a_{ij}  = \beta_{ij}+\sum_{n=i+1}^{j-1} a_{in}\beta_{nj},
\end{equation*}

\noindent for $i<j$, and zero otherwise.

Define vector $k_i=u_i'/(u_i'u_i)$ and note that $k_i'x_i=k_i'u_i=1$.

Now, equivalence in expectation between Gram-Schmidt coefficient matrix $C$ and the reduced-form true, underlying matrix A, $E[c_{ij}] = a_{ij}$, can be shown recursively, since by assumptions and Theorem 1(A) for each coefficient: \begin{align}
    E[c_{ij}]   &= E[k_i' \, x_j], \\
                &= E[k_i'(\sum_{n=1}^{j-1} x_n \, \beta_{nj} + \upsilon_j)], \\
                &= E[\sum_{n=i}^{j-1} k_i' \, x_n \, \beta_{nj} + k_i' \, \upsilon_j], \\
                \label{unbiasedresult0}
                &= \beta_{ij}+\sum_{n=i+1}^{j-1} E[c_{in}] \beta_{nj} , \\
                \label{unbiasedresult1}
                &= \beta_{ij}+\sum_{n=i+1}^{j-1} a_{in} \beta_{nj},
\end{align}

\noindent for $i<j$, and zero otherwise. Line (\ref{unbiasedresult1}) holds because $E[c_{ij}] = a_{ij}$ follows directly from (\ref{unbiasedresult0}) for $j=i+1$, proving the $j=i+2$ case, and so forth through $j=K$.

(D) Gram-Schmidt coefficient variance is lower than OLS. Define (i) $X_{-i}$ as the regressor set $X$ excluding regressor $x_i$; (ii) $X_{<i}$ as the regressor set excluding later-determined regressors $x_i,...,x_K$; (iii) $B \subseteq\mathcal{R}^{[j-1 \times 1]}$ as the OLS coefficient vector; and (iv) $C_i \subseteq\mathcal{R}^{[K \times 1]}$ as the $i^\text{th}$ Gram-Schmidt coefficient vector. Consider the non-trivial case when $\beta_{ij} \neq 0$.

\textit{\textbf{Lemma A1}}: \label{leA1}
    The coefficient of determination weakly declines as regressors are excluded: \begin{align*}
        R_{x_{i} \; X_{<i}}^2 \leq R_{x_{i} \; X_{-i}}^2.
    \end{align*}

\noindent This lemma is true for any regressor order, though for convenience, we specify order-of-exclusion in terms of regressors determined after $x_i$. The proof is standard and omitted here.

The variance of OLS coefficient $b_{ij}$ is a function of the $i^{\text{th}}$ diagonal element of the variance-covariance matrix, which can be expressed by the Schur Complement \citep[see][]{schur}, in terms of the $j^{th}$ regression coefficient of determination: \begin{align*}
    Var[b_{ij}] &= \sigma_{\upsilon_{u_j}} (X'X)_{ii}^{-1} , \\
                &= \sigma_{\upsilon_{u_j}} (x_i'x_i - B_{-i}'(X_{-i}'X_{-i})B_{-i})^{-1}, \\
                &= \sigma_{\upsilon_{u_j}} (x_i'x_i (1-R_{x_{i} \; X_{-i}}^2))^{-1}.
\end{align*}

The variance of Gram-Schmidt coefficient $c_{ij}$ is by 1(A) a function of the residual $u_i$, which by 1(A) and Lemma A1 is: \begin{align*}
    Var[c_{ij}] &= \sigma_{\upsilon_{u_j}} (u_i'u_i)^{-1} , \\
                &= \sigma_{\upsilon_{u_j}} (x_i'x_i - C_i'(X'X)C_i)^{-1}, \\
                &= \sigma_{\upsilon_{u_j}} (x_i'x_i (1-R_{x_{i} \; X_{<i}}^2))^{-1}, \\
                &\leq \sigma_{\upsilon_{u_j}} (x_i'x_i (1-R_{x_{i} \; X_{-i}}^2))^{-1}.
\end{align*}

\noindent The result holds with strict inequality for $i=1,...j-2$ and with strict equality for $i=j-1$ because the excluded regressor set is identical between the terminal Gram-Schmidt and the OLS coefficient.

\newpage
\subsection{Full Regression Results}

\begin{table}[!h] \centering 
  \caption{Total Megawatts Generating - Full Covariates} 
  \label{C} 
\tiny 
\begin{tabular}{@{\extracolsep{0pt}}lD{.}{.}{-3} D{.}{.}{-3} D{.}{.}{-3} D{.}{.}{-3} } 
\\[-1.8ex]\hline 
        \hline \\[-1.8ex]
        \multicolumn{2}{c}{}\\
        Dep. Var.: \\ Generating Cap. (MW) & \multicolumn{2}{c}{Natural Gas} & \multicolumn{2}{c}{Renewables} \\
        \cmidrule(lr){2-3}\cmidrule(lr){4-5}
        & \multicolumn{1}{c}{Direct} & \multicolumn{1}{c}{Total} & \multicolumn{1}{c}{Direct} & \multicolumn{1}{c}{Total}\\ 
        \\[-1.8ex] & \multicolumn{1}{c}{(1)} & \multicolumn{1}{c}{(2)} & \multicolumn{1}{c}{(3)} & \multicolumn{1}{c}{(4)}\\ 
        \hline \\[-1.8ex] 
2022-year & 7,842.43 & 2,846.69^{***} & 20,310.79^{***} & 25,566.32^{***} \\ 
  2021-year & 6,377.75 & 2,421.75^{***} & 16,632.67^{***} & 18,388.01^{***} \\ 
  2020-year & 5,418.43 & 2,049.98^{***} & 13,294.30^{***} & 12,068.58^{***} \\ 
  2019-year & 3,308.18 & 886.71^{***} & 9,810.73^{***} & 7,760.83^{***} \\ 
  2018-year & 2,902.03 & 618.95^{***} & 6,419.09^{***} & 4,900.03^{***} \\ 
  2017-year & 1,569.90 & 682.33^{***} & 2,737.63^{**} & 2,488.76^{***} \\ 
  12-month & 1,305.20 & 528.36^{**} & 3,079.93^{***} & 4,302.72^{***} \\ 
  11-month & 717.79 & 488.62^{**} & 2,646.77^{**} & 3,233.86^{***} \\ 
  10-month & 185.83 & 449.29^{**} & 3,141.17^{**} & 2,915.30^{***} \\ 
  9-month & -45.64 & 416.27^{*} & 3,135.82^{*} & 2,602.40^{***} \\ 
  8-month & -154.38 & 399.19^{*} & 2,793.48^{*} & 2,333.20^{***} \\ 
  7-month & -1.57 & 382.39^{*} & 2,052.31^{*} & 1,981.36^{***} \\ 
  6-month & 408.58 & 180.13 & 1,166.16 & 1,546.45^{***} \\ 
  5-month & 283.63 & -153.82 & 182.47 & 886.42^{***} \\ 
  3-month & -622.97 & -267.15 & 206.07 & 477.22^{**} \\ 
  4-month & -169.27 & -215.15 & 65.38 & 622.58^{***} \\ 
  2-month & -381.80 & -87.70 & 295.95 & 179.55 \\ 
  Midpoint Temp. - $^\circ$C 12mo mean & -202.53 & -81.44 & 77.33 & -236.32^{*} \\ 
  Midpoint Temp. Sq. - $^\circ$C  12mo Mean & 182.44 & 155.23^{*} & 171.56 & 324.48^{***} \\ 
  Mean Wind speed - km/hr 12mo mean & -2,432.79^{***} & -919.20^{***} & -956.74^{*} & -743.13^{***} \\ 
  Labor Force Pop. - Millions & 363.92 & 308.88 & 99.95 & 6,591.84^{***} \\ 
  Unemp. Rate & -47.92 & -5.49 & -22.58 & 211.92^{**} \\ 
  CPI & -82.50 & -115.02^{**} & -30.12 & 50.35 \\ 
  Interest Rate & 369.12 & 368.29 & 238.81 & -1,636.90^{***} \\ 
  Cost Wind Installation - \$/MWh & -64.63 & -42.05 & 51.45 & -264.90^{***} \\ 
  Cost PV Installation - \$/MWh & 2.22 & 37.54 & -26.43 & 136.85^{***} \\ 
  Natural Gas Price - \$/MMBtu & -436.63 & -17.21 & -322.33 & 112.17 \\ 
  Online Capacity - GW & 10.44 & 37.51 & 103.30 & -301.11^{***} \\ 
  Offline Capacity - GW & 919.01^{*} & 804.84^{**} & -617.20 & -665.78^{*} \\ 
  Capacity Utilization - \% & 475.85 & 772.47^{*} & -1,162.50^{*} & -1,023.50^{**} \\ 
  Shadow Price - \$/MW & 4.45 & 32.20^{*} & -113.66 & -30.74^{**} \\ 
  Energy Only Price - \$/MW & 12.15 & 3.78 & 65.68 & 29.44^{***} \\ 
  Incentive - \$/MW & -34.13 & -21.80 & -51.55^{*} & -24.03 \\ 
  Change Peaker Net Margin - \$/GW & 1,080.57 & 1,213.13 & 884.36 & -1,326.20 \\ 
  NG GINI Index & -72.67 & -8.10 & -191.19 & 1.70 \\ 
  Wind GINI Index & 109.22 & 165.13 & 231.34 & 302.94^{*} \\ 
  Renewable Appl. Pool - MW & -0.08 & -0.09 & 0.35^{*} & 0.30^{*} \\ 
  NG Appl. Pool - MW  & -0.18 & -0.16 & -0.59^{***} & -0.69^{***} \\ 
  Mean Status Phase - Renewables & -4.02 & -4.02 & 661.84 & 661.84 \\ 
  Mean Status Phase - NG  & -152.18 & -152.18 & 617.50 & 617.50 \\ 
  Constant & 67,125.29^{*} & 56,893.43^{***} & 29,068.73 & 28,993.32^{***} \\ 
 \hline \\[-1.8ex] 
Observations & \multicolumn{1}{c}{78} & \multicolumn{1}{c}{78} & \multicolumn{1}{c}{78} & \multicolumn{1}{c}{78} \\ 
R$^{2}$ & \multicolumn{1}{c}{0.96} & \multicolumn{1}{c}{0.96} & \multicolumn{1}{c}{1.00} & \multicolumn{1}{c}{1.00} \\ 
Adjusted R$^{2}$ & \multicolumn{1}{c}{0.92} & \multicolumn{1}{c}{0.92} & \multicolumn{1}{c}{1.00} & \multicolumn{1}{c}{1.00} \\ 
\hline 
\hline \\[-1.8ex] 
\textit{Note: $R^2$ values for Renewables}  & \multicolumn{4}{r}{$^{*}$p$<$0.05; $^{**}$p$<$0.01; $^{***}$p$<$0.001} \\ 
\textit{shown as 1 due to rounding.}  & \multicolumn{4}{r}{} \\ 
\end{tabular} 
\end{table} 

\newpage

\begin{table} 
    \centering
    \tiny
    \captionsetup{skip=0pt} 
    \captionsetup{belowskip=0pt}  
    \caption{Comparison Application Pool to Generating - Full Covariates} 
    \setlength{\tabcolsep}{0pt}
    \begin{tabular}{@{\extracolsep{1pt}}lD{.}{.}{-3} D{.}{.}{-3} D{.}{.}{-3} D{.}{.}{-3} } 
        \\[-1.8ex]\hline 
        \hline \\[-1.8ex]
        \multicolumn{2}{c}{}\\
        & \multicolumn{2}{c}{Natural Gas} & \multicolumn{2}{c}{Renewables} \\
        \cmidrule(lr){2-3}\cmidrule(lr){4-5}
        Dependent Variable: (MW) & \multicolumn{1}{c}{Appl. Pool} & \multicolumn{1}{c}{Gen. Cap.} & \multicolumn{1}{c}{Appl. Pool} & \multicolumn{1}{c}{Gen. Cap,}\\ 
        & \multicolumn{1}{c}{(1)} & \multicolumn{1}{c}{(2)} & \multicolumn{1}{c}{(3)} & \multicolumn{1}{c}{(4)}\\ 
        \hline \\[-1.8ex] 
2022-year & -4,006.19^{***} & 2,846.69^{***} & 9,548.61^{***} & 25,566.32^{***} \\ 
  2021-year & -2,340.79^{***} & 2,421.75^{***} & 4,398.01^{***} & 18,388.01^{***} \\ 
  2020-year & -117.62 & 2,049.98^{***} & 2,086.68^{***} & 12,068.58^{***} \\ 
  2019-year & 2,538.28^{***} & 886.71^{***} & 2,162.48^{***} & 7,760.83^{***} \\ 
  2018-year & 3,907.97^{***} & 618.95^{***} & 2,135.86^{***} & 4,900.03^{***} \\ 
  2017-year & 4,048.45^{***} & 682.33^{***} & 1,804.40^{***} & 2,488.76^{***} \\ 
  12-month & -703.11^{**} & 528.36^{**} & 1,612.50^{***} & 4,302.72^{***} \\ 
  11-month & -671.70^{*} & 488.62^{**} & 1,278.86^{***} & 3,233.86^{***} \\ 
  10-month & -592.84^{*} & 449.29^{**} & 968.80^{***} & 2,915.30^{***} \\ 
  9-month & -516.51 & 416.27^{*} & 844.23^{***} & 2,602.40^{***} \\ 
  8-month & -1,130.32^{***} & 399.19^{*} & 678.61^{***} & 2,333.20^{***} \\ 
  7-month & -1,571.52^{***} & 382.39^{*} & 354.18^{*} & 1,981.36^{***} \\ 
  6-month & -341.24 & 180.13 & 297.82 & 1,546.45^{***} \\ 
  5-month & -306.24 & -153.82 & 256.02 & 886.42^{***} \\ 
  3-month & -247.32 & -267.15 & -16.52 & 477.22^{**} \\ 
  4-month & -268.91 & -215.15 & 169.94 & 622.58^{***} \\ 
  2-month & -133.56 & -87.70 & -96.47 & 179.55 \\ 
  Midpoint Temp. - $^\circ$C 12mo mean & 781.47^{***} & -81.44 & 112.23 & -236.32^{*} \\ 
  Midpoint Temp. Sq. - $^\circ$C 12mo mean & 48.68 & 155.23^{*} & 291.71^{***} & 324.48^{***} \\ 
  Mean Wind speed - km/hr 12mo mean & -1,521.35^{***} & -919.20^{***} & -1,352.60^{***} & -743.13^{***} \\ 
  Labor Force Pop. - Millions & -5,302.81^{***} & 308.88 & 2,984.89^{***} & 6,591.84^{***} \\ 
  Unemp. Rate & -252.97^{*} & -5.49 & -81.58 & 211.92^{**} \\ 
  CPI & -39.47 & -115.02^{**} & 16.67 & 50.35 \\ 
  Interest Rate & 1,650.64^{***} & 368.29 & -1,330.31^{***} & -1,636.90^{***} \\ 
  Cost Wind Installation - \$/MWh & 297.06^{**} & -42.05 & 5.38 & -264.90^{***} \\ 
  Cost PV Installation - \$/MWh & -211.04^{***} & 37.54 & -22.49 & 136.85^{***} \\ 
  Natural Gas Price - \$/MMBtu & -767.11^{***} & -17.21 & 227.93 & 112.17 \\ 
  Online Capacity - GW & 619.91^{***} & 37.51 & -63.99 & -301.11^{***} \\ 
  Offline Capacity - GW & -2,053.85^{***} & 804.84^{**} & -351.79 & -665.78^{*} \\ 
  Capacity Utilization - \% & 368.73 & 772.47^{*} & -217.99 & -1,023.50^{**} \\ 
  Shadow Price - \$/MW & 40.73^{*} & 32.20^{*} & -33.97^{**} & -30.74^{**} \\ 
  Energy Only Price - \$/MW & -1.90 & 3.78 & 15.53^{*} & 29.44^{***} \\ 
  Incentive - \$/MW & -31.71 & -21.80 & 19.31 & -24.03 \\ 
  Change Peaker Net Margin - \$/GW & -5,397.12^{***} & 1,213.13 & -2,601.29^{**} & -1,326.20 \\ 
  NG GINI Index & 136.93^{***} & -8.10 & 26.88^{***} & 1.70 \\ 
  Wind GINI Index & -190.87 & 165.13 & -72.43 & 302.94^{*} \\ 
  Renewable Appl. Pool - MW &  & -0.09 &  & 0.30^{*} \\ 
  NG Appl. Pool - MW  &  & -0.16 &  & -0.69^{***} \\ 
  Mean Status Phase - Renewables &  & -4.02 &  & 661.84 \\ 
  Mean Status Phase - NG  &  & -152.18 &  & 617.50 \\ 
  Constant & 7,080.37^{***} & 56,893.43^{***} & 4,699.32^{***} & 28,993.32^{***} \\ 
 \hline \\[-1.8ex] 
Observations & \multicolumn{1}{c}{78} & \multicolumn{1}{c}{78} & \multicolumn{1}{c}{78} & \multicolumn{1}{c}{78} \\ 
R$^{2}$ & \multicolumn{1}{c}{0.99} & \multicolumn{1}{c}{0.96} & \multicolumn{1}{c}{0.99} & \multicolumn{1}{c}{1.00} \\ 
Adjusted R$^{2}$ & \multicolumn{1}{c}{0.98} & \multicolumn{1}{c}{0.92} & \multicolumn{1}{c}{0.99} & \multicolumn{1}{c}{1.00} \\ 
\hline 
\hline \\[-1.8ex] 
\textit{Note: $R^2$ values for Renewables}  & \multicolumn{4}{r}{$^{*}$p$<$0.05; $^{**}$p$<$0.01; $^{***}$p$<$0.001} \\ 
\textit{shown as 1 due to rounding.}  & \multicolumn{4}{r}{} \\ 
\end{tabular} 
\end{table} 

\newpage
\subsection{\normalsize{Additional Robustness Checks}}

\begin{center}
{\textit{\TenPtType{A.5.1 Autocorrelation}}}
\end{center}

When autocorrelation is present, coefficient estimates remain unbiased but no longer achieve minimum variance. To test for the presence and impact of autocorrelation, we run a Generalized Least Squares model \citep{Goldberger1962} with a lagged error term of one-period (AR1) and 10-periods (AR10). Results are compared in Tables \ref{Tbl_AR1underFullerrors}, \ref{Tbl_AR10underFullerrors},  and \ref{Tbl_underComparison}. Positive first-order autocorrelation is detected with a coefficient for the lagged error of 0.86 and 0.67 in the two models, respectively, both significant at the 0.1\% confidence level, suggesting trader pricing errors are persistent. The incentive coefficient of interest remains negative and significant across all models.

\begin{table}[h]
\caption{AR1 Underbidding Direct and Total Effects - Omitting FE} 
  \label{Tbl_AR1underFullerrors} 
\small
\begin{tabular}{@{\extracolsep{1pt}}lD{.}{.}{-3} D{.}{.}{-3} } 
\\[-1.8ex]\hline 
\hline \\[-1.8ex] 
{\textit{Dependent variable:}} & \multicolumn{2}{c}{Energy Price} \\
\cline{2-3}
& \multicolumn{1}{c}{Direct} & \multicolumn{1}{c}{Total}\\ 
 & \multicolumn{1}{c}{(1)} & \multicolumn{1}{c}{(2)}\\ 
\hline \\[-1.8ex] 
  Nat. Gas Capacity - GW & -10.71^{***} & -24.51^{***} \\ 
  Renewable Capacity - GW & -21.04^{***} & -24.28^{***} \\ 
  Other Capacity - GW & -19.41^{***} & 67.87^{***} \\ 
  Midpoint Temp. - $^\circ$C  & -2.38^{***} & -2.15^{***} \\ 
  Midpoint Temp. Sq. - $^\circ$C & 0.68^{***} & 0.76^{***} \\ 
  Wind speed - Km/hr & -1.17^{***} & -1.33^{***} \\ 
  Natural Gas Price - \$/MMBtu & 36.21^{***} & 31.01^{***} \\ 
  15min AR Error & 0.86^{***} & 0.86^{***} \\ 
  Supply Nat. Gas - GW & -51.02^{***} & 16.14^{***} \\ 
  Supply Renewables - GW & -44.10^{***} & 0.68^{***} \\ 
  Supply Other - GW & -93.28^{***} & -15.24^{***} \\ 
  Reserve Capacity - GW & 30.46^{***} & -7.09^{***} \\ 
  Capacity Utilization - \% & 25.65^{***} & 22.86^{***} \\ 
  Incentive Active - Y/N & -14.07^{***} & -13.31^{***} \\ 
  Incentive Payment - \$/MW & -0.09^{***} & -0.09^{***} \\ 
  Constant & 441.55^{***} & 30.70^{***} \\ 
 \hline \\[-1.8ex] 
Observations & \multicolumn{1}{c}{257,251} & \multicolumn{1}{c}{257,251} \\ 
R$^{2}$ & \multicolumn{1}{c}{0.82} & \multicolumn{1}{c}{0.82} \\ 
Adjusted R$^{2}$ & \multicolumn{1}{c}{0.82} & \multicolumn{1}{c}{0.82} \\
\hline \\[-1.8ex] 
\hline \\
\textit{Note:Excludes Fixed Effects}  & \multicolumn{2}{r}{$^{*}$p$<$0.05; $^{**}$p$<$0.01; $^{***}$p$<$0.001} \\ 
\end{tabular} 
\end{table} 

\newpage

\begin{table}[!htbp] \centering 
\caption{AR10 Underbidding Direct and Total Effects - Omitting FE} 
  \label{Tbl_AR10underFullerrors} 
\small 
\begin{tabular}{@{\extracolsep{1pt}}lD{.}{.}{-2} D{.}{.}{-2} } 
\\[-1.8ex]\hline 
\hline \\[-1.8ex] 
 & \multicolumn{2}{c}{\textit{Dependent variable:}} \\ 
\cline{2-3} 
\\[-1.8ex] & \multicolumn{2}{c}{Energy Price} \\ 
\\[-1.8ex] & \multicolumn{1}{c}{(1)} & \multicolumn{1}{c}{(2)}\\ 
\hline \\[-1.8ex] 
  Nat. Gas Capacity - GW  & -11.54^{***} & -24.52^{***} \\ 
  Renewables Capacity - GW & -22.78^{***} & -24.28^{***} \\ 
  Other Capacity - GW & -19.02^{***} & 67.89^{***} \\ 
  120hr AR Error & -0.01^{***} & 0.01^{***} \\ 
  96hr AR Error & 0.02^{***} & 0.06^{***} \\ 
  72hr AR Error & 0.03^{***} & 0.09^{***} \\ 
  48hr AR Error & 0.04^{***} & 0.32^{***} \\ 
  24hr AR Error & 0.04^{***} & 0.37^{***} \\ 
  Midpoint Temp. - $^\circ$C & -2.49^{***} & -2.29^{***} \\ 
  Midpoint Temp. Sq. - $^\circ$C & 0.68^{***} & 0.67^{***} \\ 
  Wind speed - Km/hr & -1.29^{***} & -1.34^{***} \\ 
  Natural Gas Price - \$/MMBtu & 37.80^{***} & 25.90^{***} \\ 
  75min AR Error & 0.04^{***} & 0.59^{***} \\ 
  60min AR Error & 0.09^{***} & 0.45^{***} \\ 
  45min AR Error & -0.01^{***} & 0.44^{***} \\ 
  30min AR Error & 0.10^{***} & 0.57^{***} \\ 
  15min AR Error & 0.67^{***} & 0.69^{***} \\ 
  Supply Nat. Gas - GW & -50.47^{***} & 16.37^{***} \\ 
  Supply Renewables - GW & -43.58^{***} & 1.02^{***} \\ 
  Supply Other - GW & -93.33^{***} & -14.91^{***} \\ 
  Reserve Capacity - GW & 30.34^{***} & -6.86^{***} \\ 
  Capacity Utilization - \% & 25.48^{***} & 22.35^{***} \\ 
  Incentive Active - Y/N & -12.00^{***} & -10.90^{***} \\ 
  Incentive Payment - \$/MW & -0.13^{***} & -0.13^{***} \\ 
  Constant & 503.15^{***} & 30.72^{***} \\ 
 \hline \\[-1.8ex] 
Observations & \multicolumn{1}{c}{256,772} & \multicolumn{1}{c}{256,772} \\ 
R$^{2}$ & \multicolumn{1}{c}{0.83} & \multicolumn{1}{c}{0.83} \\ 
Adjusted R$^{2}$ & \multicolumn{1}{c}{0.83} & \multicolumn{1}{c}{0.83} \\ 
\hline 
\hline \\[-1.8ex] 
\textit{Note:}  & \multicolumn{2}{r}{$^{*}$p$<$0.05; $^{**}$p$<$0.01; $^{***}$p$<$0.001} \\ 
\end{tabular} 
\end{table} 

\newpage

\begin{center}
{\textit{\TenPtType{A.5.2 Extreme Events and Covariate Exclusion}}}
\end{center}

To test for the influence of omitted variables and extreme events on Generating Capacity, covariate groups and extreme event observations are excluded sequentially in the table below. Covariates are grouped by Fixed Effects (annual), Seasonal Effects (monthly), climatic effects (temperature and wind), economic effects (Labor Force, Unemp. Rate, CPI, Interest Rate, Cost Wind \& PV Installation, Natural Gas Price), Upstream Market Effects (Online \& Offline Capacity, Capacity Utilization, Shadow Price, Energy Only Price) and Downstream Market Effects (Compensation Adder, Change Pekaer Net Margin, NG \& Renewable GINI Index, Renewable \& NG Applicant Pool, and Renewable \& NG Status Phase). Results are shown for 12, 24, and 36-month lagged models. Exclusion of Winter Storm Uri (Feb. 2021) is also shown. 

\begin{table}
\centering 
\captionsetup{font=small}
  \caption{Covariate Sequencing - Generating Capacity Incentive Effect: 12 Month Lag}
  \label{D} 
\tiny 
\begin{tabular}{@{\extracolsep{5pt}} lccccccc} 
\\[-1.8ex]\hline 
\hline \\[-1.8ex] 
 & (1) & (2) & (3) & (4) & (5) & (6) & (7) \\ 
\hline \\[-1.8ex] 
Incentive & -0.971 & 0.175 & 1.122 & 1.115 & -21.796 & -34.131 & -30.208 \\ 
\hline \\[-1.8ex] 
Fixed Effects & X & X & X & X & X & X & X \\ 
Seasonal Effects &  & X & X & X & X & X & X \\ 
Climatic Eff. &  &  & X & X & X & X & X \\ 
Economic Eff. &  &  &  & X & X & X & X \\ 
Upstream Market Eff. &  &  &  &  & X & X & X \\ 
Downstream Market Eff. &  &  &  &  &  & X & X \\ 
Exclude Storm Uri &  &  &  &  &  &  & X \\ 
\hline \\[-1.8ex] 
Obs & 78 & 78 & 78 & 78 & 78 & 78 & 77 \\ 
R2 & 0.822 & 0.889 & 0.918 & 0.935 & 0.956 & 0.963 & 0.962 \\ 
AIC & 1190.462 & 1175.691 & 1158.004 & 1153.788 & 1134.030 & 1134.050 & 1123.497 \\ 
BIC & 1211.672 & 1222.825 & 1212.209 & 1224.489 & 1216.515 & 1233.032 & 1221.937 \\ 
F Statistic & 46.141 & 26.207 & 29.832 & 25.188 & 28.724 & 24.021 & 22.706 \\ 
\hline \\[-1.8ex] 
\end{tabular} 
\end{table} 

\begin{table}[!htbp] \centering 
\captionsetup{font=small}
  \caption{Covariate Sequencing - Generating Capacity Incentive Effect: 24 Month Lag}
  \label{E} 
\tiny 
\begin{tabular}{@{\extracolsep{5pt}} lccccccc} 
\\[-1.8ex]\hline 
\hline \\[-1.8ex] 
 & (1) & (2) & (3) & (4) & (5) & (6) & (7) \\ 
\hline \\[-1.8ex] 
Incentive & -0.973 & -0.514 & -0.523 & -1.1\textasteriskcentered  & 16.109\textasteriskcentered  & 5.145 & -0.684 \\ 
\hline \\[-1.8ex] 
Fixed Effects & X & X & X & X & X & X & X \\ 
Seasonal Effects &  & X & X & X & X & X & X \\ 
Climatic Eff. &  &  & X & X & X & X & X \\ 
Economic Eff. &  &  &  & X & X & X & X \\ 
Upstream Market Eff. &  &  &  &  & X & X & X \\ 
Downstream Market Eff. &  &  &  &  &  & X & X \\ 
Exclude Storm Uri &  &  &  &  &  &  & X \\ 
\hline \\[-1.8ex] 
Obs & 66 & 66 & 66 & 66 & 66 & 66 & 65 \\ 
R2 & 0.914 & 0.946 & 0.959 & 0.976 & 0.985 & 0.991 & 0.992 \\ 
AIC & 934.806 & 926.979 & 914.337 & 892.913 & 873.106 & 853.943 & 836.642 \\ 
BIC & 952.323 & 968.582 & 962.510 & 956.413 & 947.555 & 943.719 & 925.792 \\ 
F Statistic & 105.168 &  49.064 &  52.593 &  57.268 &  66.504 &  71.494 &  76.080 \\ 
\hline \\[-1.8ex] 
\end{tabular} 
\end{table} 

\begin{table}[!htbp] \centering
\captionsetup{font=small}
  \caption{Covariate Sequencing - Generating Capacity Incentive Effect: 36 Month Lag}
  \label{F} 
\tiny 
\begin{tabular}{@{\extracolsep{5pt}} lccccccc} 
\\[-1.8ex]\hline 
\hline \\[-1.8ex] 
 & (1) & (2) & (3) & (4) & (5) & (6) & (7) \\ 
\hline \\[-1.8ex] 
Incentive & -0.973\textasteriskcentered  & -0.612 & -0.704\textasteriskcentered  & -0.687\textasteriskcentered  & -1.414 & -8.819 & -4.895 \\
\hline \\[-1.8ex] 
Fixed Effects & X & X & X & X & X & X & X \\ 
Seasonal Effects &  & X & X & X & X & X & X \\ 
Climatic Eff. &  &  & X & X & X & X & X \\ 
Economic Eff. &  &  &  & X & X & X & X \\ 
Upstream Market Eff. &  &  &  &  & X & X & X \\ 
Downstream Market Eff. &  &  &  &  &  & X & X \\ 
Exclude Storm Uri &  &  &  &  &  &  & X \\ 
\hline \\[-1.8ex] 
Obs & 54 & 54 & 54 & 54 & 54 & 54 & 53 \\ 
R2 & 0.926 & 0.971 & 0.981 & 0.996 & 0.996 & 0.998 & 0.998 \\ 
AIC & 753.927 & 724.613 & 707.436 & 640.688 & 644.292 & 634.977 & 625.207 \\ 
BIC & 767.850 & 760.415 & 749.205 & 696.379 & 709.929 & 714.536 & 704.019 \\ 
F Statistic & 119.692 &  78.239 &  93.956 & 246.819 & 189.972 & 162.935 & 153.457 \\ 
\hline \\[-1.8ex] 
\end{tabular} 
\end{table} 

To test the influence of omitted variables and outlier observations on the Direct Underbidding effect and the Average Treatment Effect, covariate groups are added, and rare event observations are removed sequentially in the table below. Columns (1)-(3) compare Control covariates excluding rare events, including observations occurring during Storm Uri, and those when the total energy price, including incentives, is within \$2,000 of the regulatory cap. Columns (4)-(5) examine the layering of one-period (AR1) and 10-period (AR10) lagged error covariates excluding regulatory cap observations. Columns (6)-(8) report the average total incentive for covariate matching when excluding Storm Uri observations(6),  regulatory cap observations (7), and with the inclusion of AR1 and AR10 covariates while excluding regulatory cap observations(8). The Average Active Incentive per megawatt is included for each data subset for comparison.

\begin{table}[!h] \centering
    \captionsetup{font=small}
  \caption{Underbidding Incentive Effect - Direct and Treatment Effect Robustness} 
  \label{Tbl_underComparison} 
\tiny 
\addtolength{\tabcolsep}{-0.4em}
\begin{tabular}{@{\extracolsep{0pt}} lcccccccc} 
\\[-1.8ex]\hline 
\hline \\[-1.8ex]
 & \multicolumn{5}{c}{Direct} & \multicolumn{3}{c}{ATET}  \\
  \cmidrule(lr){2-6}\cmidrule(lr){7-9} \\
Dep.: Energy Price - \$/MW & (1) & (2) & (3) & (4) & (5) & (6) & (7) & (8) \\ 
\\[-1.8ex] 
\hline \\[-1.8ex] 
Active Incentive Effect &  -79.96$^{***}$&  -15.14$^{***}$     &   -15.19$^{***}$ &  -5.27$^{***}$ & -5.78$^{***}$  &  -224.78$^{***}$ & -390.98$^{***}$   & -24.91$^{***}$\\
Incentive Effect & -0.05$^{***}$&  0.24$^{***}$  &   0.07$^{***}$ & -0.21$^{***}$ &  -0.26$^{***}$  & - & -  & - \\
\hline \\[-1.8ex]
Avg. Active Incentive -\$/MW & 55.31 & 12.20& 31.72&12.20 &12.20  & 12.20 & 31.72  & 12.20 \\
Avg. Underbid -\$/MW & 79.73& 12.21 & 12.97 & 7.83 & 8.95 & 224.78 & 390.98  & 24.91 \\
\hline \\[-1.8ex] 
FE and Controls & X & X & X & X & X & X & X & X\\ 
AR1 &  &  & & X & X &  &   & X\\ 
AR10 &  &  &  & & X &  &   & X \\ 
Matching &  &  &   &  &  & X & X & X \\ 
Excl. Tot. Payment $>$\$7,000 &  & X & & X & X & X &  & X\\
Exclude Uri &  &  & X &  & &  & X & \\ 
\hline \\[-1.8ex] 
Obs &  257,252         &    256,850            &  256,484 & 256,849 & 256,370  &  101,620 & 101,220   & 101,460\\ 
R2 &  0.312         &     0.170         &     0.132  &  0.714 & 0.732    & - &  -  & -\\ 
AIC & 3,281,245        &  3,006,953         & 2,926,943 & 2,732,782 & 2711664     & - & - & -\\ 
BIC & 3,282,144         &  3,007,851          &  2,927,843 & 2,733,692 & 2712668 & - & -   &-\\ 
F Statistic & 1,373         &  617           &  458     & 7,469 & 7,377  &  - & -  &-\\ 
\hline \\[-1.8ex] 
\end{tabular} 
\end{table} 

\begin{center}
    \begin{figure}[!h]
        \includegraphics[height=6cm]{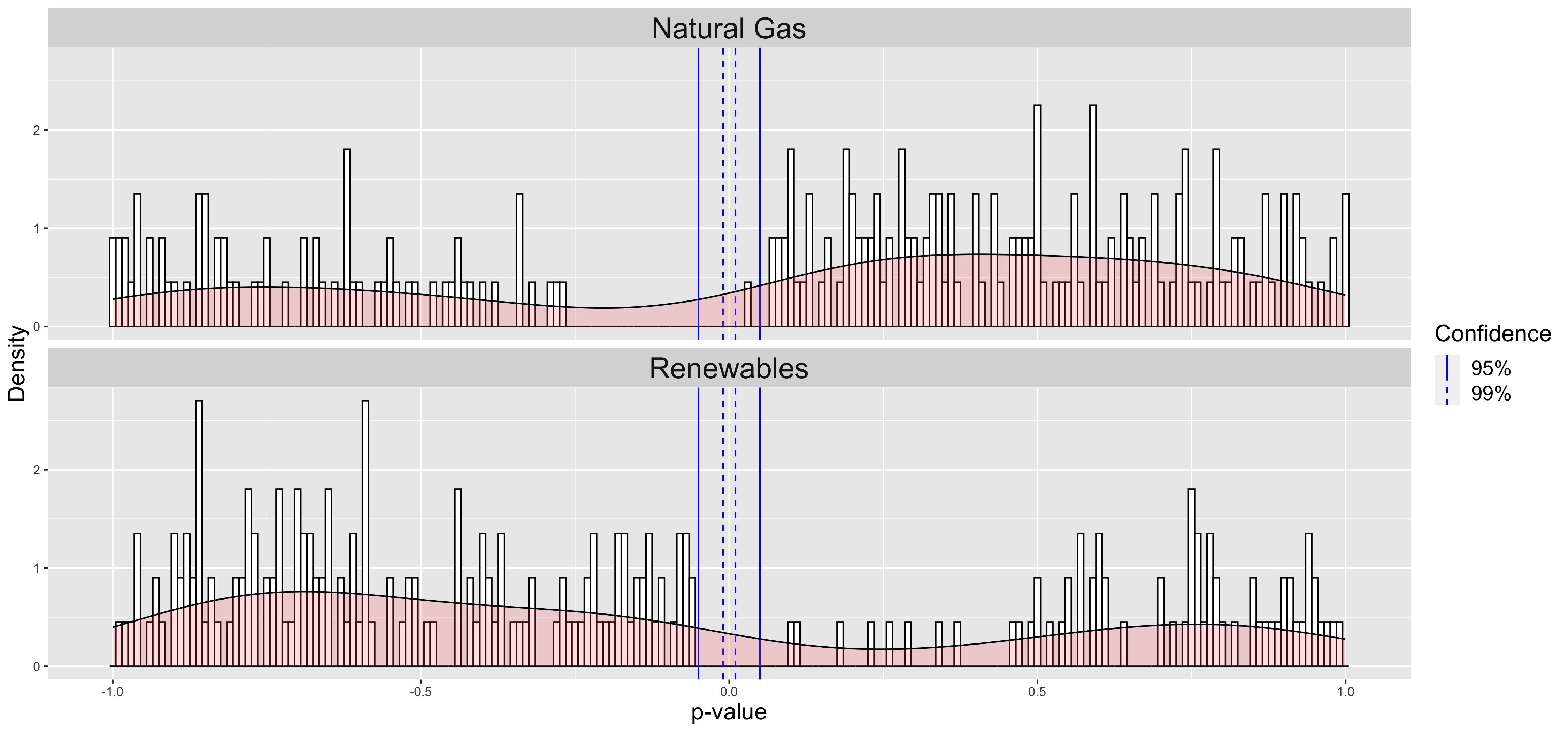}
        \caption{Robustness Results: Incentive Significance and Direction}
        \label{fig:lags_robustness}
        \begin{figurenotes}[Note]
        Negative p-values indicate negative Incentive coefficient estimates. 
        \end{figurenotes}
    \end{figure}
\end{center}

\begin{center}
{\textit{\TenPtType{A.5.3 Information Lags and Covariate Exclusion}}}
\end{center}
Figure \ref{fig:lags_robustness} displays a histogram and density plot of coefficient p-values for 222 model alternatives for both Natural Gas and Renewable Capacity shown in Table \ref{TBL_GEN}. Covariates lags range from zero to 36 months. For each lag, three fixed effects permutations are evaluated: no fixed effects, annual fixed effects, and both annual and seasonal fixed effects. Each version is repeated with and without a polynomial expansion on temperature. The Incentive coefficient of interest was found to be both positive and significant at the 95\% confidence level in one model, or 0.45\% of those evaluated. No Renewable Capacity models were significant at the 95\% confidence level or above.

\newpage

\begin{center}
{\textit{\TenPtType{A.5.4 Rebalancing}}}
\end{center}

The covariate matching method in of Table \ref{TBL_match} resulted in the following observation counts.

\begin{table}[h]
    \centering
    \caption{Underbidding - Covariate Matching Observation Count}
    \small
    \begin{tabular}{l*{2}{c}}
    \hline\hline
                &         Raw&     Matched\\
    \hline
    Number of obs&      256,772&      102,152\\
Treated obs &       51,076&       51,076\\
Control obs &      205,696&       51,076\\
    \hline\hline
    \end{tabular}
\end{table}

\end{document}